%% file: main.tex
\pdfoutput=1
\documentclass{article}
\usepackage{fancyhdr}
\PassOptionsToPackage{numbers,compress}{natbib}

\usepackage{report}

\usepackage{soul}
\usepackage[utf8]{inputenc} 
\usepackage[T1]{fontenc}    
\usepackage{hyperref}       
\usepackage{url}            
\usepackage{booktabs}       
\usepackage{amsfonts}       
\usepackage{nicefrac}       
\usepackage{microtype}      
\usepackage{graphicx}
\usepackage{pifont}
\usepackage{multirow}
\usepackage{float}
\usepackage{subcaption}
\usepackage{placeins}
\usepackage{tcolorbox}
\usepackage{amsmath}
\usepackage{amssymb}
\usepackage{xspace}
\usepackage{enumitem}
\usepackage{colortbl}
\usepackage{array}
\usepackage{caption}
\usepackage{makecell}
\usepackage{listings}
\tcbuselibrary{breakable}

\lstdefinestyle{promptstyle}{
    basicstyle=\ttfamily\scriptsize,
    breaklines=true,
    breakatwhitespace=false,
    showstringspaces=false,
    columns=fullflexible,
    frame=single,
    framesep=4pt,
    rulecolor=\color{black!40},
    backgroundcolor=\color{boxgray},
    xleftmargin=4pt,
    xrightmargin=4pt,
    aboveskip=6pt,
    belowskip=6pt,
    upquote=true
}

\definecolor{groupblue}{HTML}{F5D2B0}
\definecolor{groupgreen}{HTML}{F4EBC5}
\definecolor{rowcream}{HTML}{FCF8F0}
\definecolor{highlight}{HTML}{C8E2C2}
\definecolor{boxgray}{HTML}{F2F2F2}
\definecolor{crossred}{HTML}{DC3030}
\definecolor{checkgreen}{HTML}{1FA855}
\definecolor{headerblue}{HTML}{1F4E79}
\definecolor{tagblue}{HTML}{5694D4}
\definecolor{tagred}{HTML}{E26A66}

\definecolor{darkmagenta}{rgb}{0.56, 0.0, 1.0}
\definecolor{softyellow}{rgb}{1.0, 0.92, 0.3}
\definecolor{LightAquamarine}{rgb}{0.75, 1.0, 0.8}
\definecolor{FireBrick}{RGB}{178,34,34}
\definecolor{MediumPurple}{RGB}{147,112,219}
\definecolor{uclablue}{rgb}{0.15, 0.45, 0.68}

\definecolor{njuPurple}{RGB}{220,205,230}
\definecolor{njuPurpleLight}{RGB}{250,245,252}

\newtcolorbox{abstractbox}{
    colback=njuPurpleLight,
    colframe=njuPurple,
    boxrule=1pt,
    arc=4mm,
    left=8pt,
    right=8pt,
    top=8pt,
    bottom=8pt,
    opacityback=0.95
}

\title{AVE-Compass: Towards Holistic Evaluation for Audio-Video Editing Abilities}

\author{
Yuqing Wen\textsuperscript{3,*}\quad
Yukai Huang\textsuperscript{4,*}\quad
Qianqian Xie\textsuperscript{1,*}\quad
Jiangtao Wu\textsuperscript{1}\quad
Yibin Lin\textsuperscript{6}\\
Yikai Gu\textsuperscript{5}\quad
Jialu Chen\textsuperscript{2}\quad
Yuanxing Zhang\textsuperscript{2}\quad
Jiaheng Liu\textsuperscript{1,$\dagger$}\\[6pt]
\textsuperscript{1} NJU-LINK Team, Nanjing University\quad
\textsuperscript{2} Kling Team, Kuaishou Technology\\
\textsuperscript{3} National University of Singapore\quad
\textsuperscript{4} Beijing University of Posts and Telecommunications\quad
\textsuperscript{5} University of Illinois Urbana-Champaign\quad
\textsuperscript{6} Xi'an Jiaotong University\\[4pt]
\texttt{e1351400@u.nus.edu}\quad
\texttt{liujiaheng@nju.edu.cn}
}

\begin{document}

\maketitle
\let\oldthefootnote\thefootnote
\let\thefootnote\relax\footnotetext{*~Equal Contribution. ~~$^\dagger$~Corresponding Author.}
\let\thefootnote\oldthefootnote

\input{content/0_Abstract}

\input{content/1_Introduction}
\input{content/2_RelatedWork}
\input{content/3_Benchmark}
\input{content/4_Agent}
\input{content/5_Experiments}
\input{content/6_Conclusion}

\bibliographystyle{unsrtnat}
\bibliography{references}

\clearpage
\appendix
\input{content/Appendix_Benchmark}
\input{content/Appendix_Agent}

\input{content/Appendix_Experiment}
\input{content/Appendix_BroaderImpact}
\input{content/Appendix_Limitations}
\input{content/Appendix_Prompt}

\end{document}

%% file: content/0_Abstract.tex
\begin{abstractbox}
\begin{center}
\textbf{\Large Abstract}
\end{center}
While instruction-based video editing has advanced rapidly, real-world videos contain tightly coupled audio and visual signals, and editing one modality often requires coordinated changes in the other. Existing benchmarks primarily evaluate visual transformations on silent clips or isolated audio editing, leaving complex audio-visual editing and cross-modal consistency underexplored. We introduce \textbf{AVE-Compass}\textsuperscript{\textit{a,b}}, a comprehensive benchmark with 145 curated source videos, 196 audio-visually coupled editing instructions, and 2,688 fine-grained checklist items. It evaluates Instruction Following, Fidelity Preserving, Realism, and Editing Intent through checklist-based MLLM judging and a dedicated realism rubric, complemented by automated cross-modal, video, and audio metrics. Extensive evaluation shows that state-of-the-art models still struggle to execute cross-modal instructions while preserving non-target content. We further propose \textbf{AVE-Agent}, a modular agent framework that decomposes complex instructions into dependent subtasks and iteratively improves editing results through self-reflection and evaluator feedback. AVE-Agent improves instruction execution, Fidelity Preserving, and audio-visual alignment in joint editing while maintaining competitive perceptual quality.
\par\vspace{-1mm}
\noindent\rule{0.32\linewidth}{0.4pt}\\[-1mm]
{\footnotesize
\textsuperscript{\textit{a}}\url{https://github.com/NJU-LINK/AVE-Compass}\\
\textsuperscript{\textit{b}}\url{https://huggingface.co/datasets/NJU-LINK/AVE-Compass}}
\end{abstractbox}

%% file: content/1_Introduction.tex
\section{Introduction}
\label{sec:introduction}

Recently, instruction-guided video editing models have made significant progress, enabling users to modify subjects, actions, styles, and scenes in videos through natural language instructions~\citep{qi2023fatezero,geyer2023tokenflow,kara2024rave,ku2024anyv2v,wang2025wan,jiang2025vace}. However, real-world videos are not silent visual sequences, as their semantics emerge from visual dynamics together with foreground sounds, background ambience, and speech content~\citep{chen2020vggsound,cheng2025mmaudio,cao2026t2avcompass}.
Realistic editing requests often require synchronized modifications across audio and vision. For example, a model may need to change the corresponding sound when a visible action is modified, remove the acoustic trace when a visual target is removed, or preserve lip motion and background ambience when speech-related content is edited. 
Handling such instructions requires models to understand implicit cross-modal requirements, maintain audio-visual synchronization, and avoid unintended modifications to non-target visual regions and non-target sounds~\citep{ishii2025coherent,xu2025schrodinger,lin2026zero}.

However, existing benchmarks are insufficient for evaluating this capability in both task design and evaluation protocol. Most video editing benchmarks focus on visual transformations of silent videos, assessing whether models can modify visual attributes, objects, actions, or styles~\citep{sun2025ve,chen2025editboard,chen2025ivebench,li2025five,wang2025tdve,gao2026vefx,wu2025veditbench}. Audio-related evaluations, in contrast, usually focus on isolated audio generation, audio replacement, or single synchronization scenarios~\citep{zheng2025audio,feng2025unisync}. More importantly, existing metrics are too coarse-grained to diagnose the failures that truly matter in audio-visual editing~\citep{xu2021videoclip,elizalde2023clap,huang2024vbench,goncalves2024perceptual}. 
Therefore, a benchmark for realistic audio-visual editing should evaluate not only whether the specified edit is completed, but also whether unedited content is faithfully preserved and whether the edited audio and visual signals remain physically and temporally consistent.

To address these problems, we introduce \textbf{AVE-Compass}, a comprehensive benchmark for free-form audio-visual editing. AVE-Compass comprises 145 carefully curated source videos and 196 audio-visually coupled editing instructions, together with 2,688 fine-grained checklist items. These instructions cover diverse real-world editing scenarios, ranging from local visual or acoustic modifications to joint audio-visual reconstruction, with a particular focus on models' ability to handle cross-modal dependencies and implicit synchronization constraints. To enable diagnostic evaluation, AVE-Compass decouples model performance into Instruction Following, Fidelity Preserving, Realism, and Editing Intent. Specifically, MLLM-based checklists judge edit execution and non-target content preservation, while a separate MLLM rubric evaluates perceptual realism~\citep{liu2023g,zheng2023judging}; automated metrics further measure cross-modal, video, and audio quality signals.

Systematic evaluations with AVE-Compass show that current open-source and closed-source systems are still far from reliable audio-visual editing. The most prominent failures occur on the audio side. For example, models tend to incorrectly remove or distort background sounds, or generate audio that is temporally inconsistent with the edited visual events. We argue that existing models lack an explicit mechanism for planning editing operations across different modalities while jointly enforcing fidelity preserving and synchronization constraints.
Motivated by these findings, we propose \textbf{AVE-Agent}, a planning-based agent framework for audio-visual editing. In this framework, we decompose free-form instructions into task topology graphs and use a self-reflection mechanism to inspect intermediate editing states, thereby improving generation quality and audio-visual alignment in complex joint editing scenarios~\citep{yao2022react,shen2023hugginggpt,schick2023toolformer,madaan2023self}.

\begin{figure}[t] 
    \centering
    \includegraphics[width=1.0\linewidth]{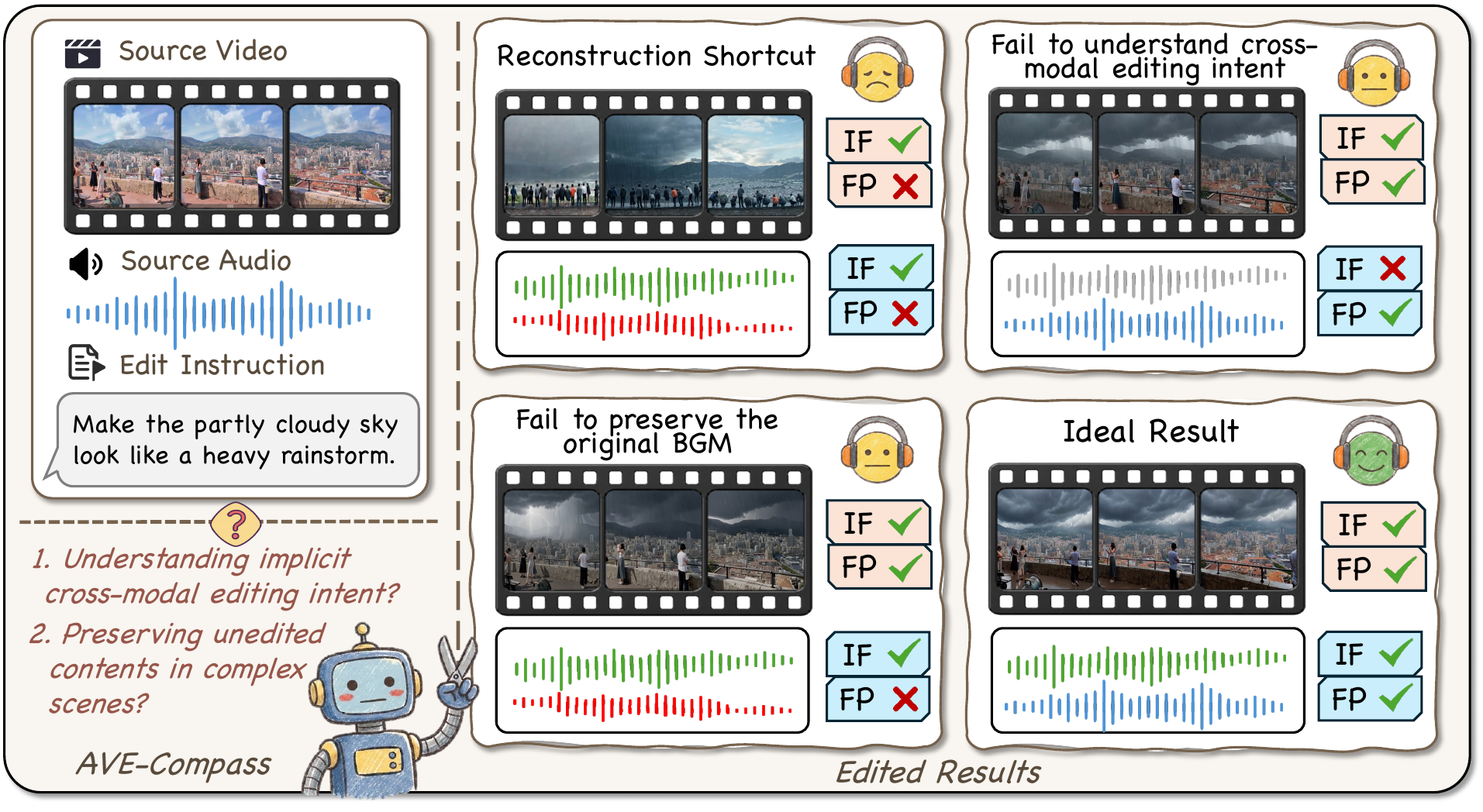}
    \caption{\textbf{Illustration of AVE-Compass.} Given the same editing instruction, the four panels illustrate characteristic ways current audio-visual editing models drift, with IF (Instruction Following) and FP (Fidelity Preserving) marks indicating where each output breaks.}
    \label{fig:intro}
\end{figure}

The main contributions of this work are summarized as follows:
\begin{itemize}[leftmargin=*, itemsep=2pt, topsep=2pt, parsep=0pt]
\item We propose \textbf{AVE-Compass}, a comprehensive benchmark for free-form audio-visual editing. It contains 145 source videos, 196 audio-visually coupled editing instructions, and 2,688 fine-grained checklist items, targeting realistic scenarios that require synchronized cross-modal editing and strict preservation of non-target content.

\item We design a decoupled evaluation protocol covering Instruction Following, Fidelity Preserving, Realism, and Editing Intent.  With these signals, AVE-Compass can diagnose incomplete edit execution, non-target content drift, low perceptual realism, and audio-visual misalignment.

\item We evaluate representative open-source and closed-source editing systems, revealing persistent failures under cross-modal instructions. We further propose \textbf{AVE-Agent}, a planning-based audio-visual editing agent framework. Through task topology decomposition and self-reflection, AVE-Agent improves editing quality in complex joint editing tasks.
\end{itemize}

%% file: content/2_RelatedWork.tex
\section{Related Work}
\label{sec:related}

\textbf{Audio-Visual Editing Methods.}
Video editing methods have recently evolved from diffusion-based zero-shot visual editing to more flexible instruction-guided editing and unified video generation/editing frameworks. Early methods typically leverage pre-trained image generation models and extend image editing capabilities to videos through attention injection, feature propagation, or temporal consistency constraints~\citep{qi2023fatezero,ceylan2023pix2video,geyer2023tokenflow,kara2024rave}. Later, instruction-guided methods further reduce the user burden, enabling models to perform more open-ended video modifications from natural language instructions~\citep{ku2024anyv2v,wu2025insvie,jiang2025vace}. Meanwhile, audio generation and editing methods have also advanced in text-to-audio generation, video-conditioned audio generation, and instruction-based audio editing~\citep{liu2023audioldm,cheng2025mmaudio,wang2023audit,lansmartdj}. However, these methods remain largely centered on a single modality. Visual editing often treats audio as a post-processing component, while audio editing lacks explicit constraints from visual events, action timing, and scene semantics. Recent audio-visual editing methods have begun to explore cross-modal consistency. They jointly modify audio and visual content through sounding-object alignment, instance-level editing, or audio-visual synchronization mechanisms~\citep{lin2026zero,fu2025object,xu2025schrodinger,zheng2025audio}. However, their task scopes are usually concentrated on specific objects, instance-level operations, or sound-effect modifications. In contrast, free-form instructions in real-world creation often involve visual, audio, and speech-related requirements simultaneously. Motivated by this need, AVE-Agent decomposes natural-language instructions into executable cross-modal subtasks and explicitly checks synchronization constraints during editing.

\textbf{Audio-Visual Editing Benchmarks.}
Existing evaluation efforts have advanced editing models from different perspectives. Visual editing benchmarks mainly evaluate whether models can modify subjects and attributes, as well as actions, styles, and visual effects in silent videos. They have also gradually introduced multi-dimensional quality assessment and fine-grained object-level control~\citep{sun2025ve,chen2025editboard,chen2025ivebench,li2025five,wang2025tdve,gao2026vefx,wu2025veditbench}. Audio evaluation, in contrast, focuses on instruction execution, subjective quality, and the preservation of audio characteristics before and after editing, providing more systematic evaluation signals for audio editing~\citep{jia2025towards}. However, these two lines of evaluation usually treat vision and audio separately, making it difficult to measure how a single editing operation is coordinated across both modalities. Furthermore, existing global similarity, quality, or synchronization metrics can reflect certain aspects of generation quality or audio-visual correspondence~\citep{xu2021videoclip,elizalde2023clap,huang2024vbench,feng2025unisync}, but they still struggle to determine whether a model understands the implicit cross-modal requirements in an instruction. For example, they cannot reliably tell whether the corresponding sound changes after an action is modified, whether background sounds are preserved after a visual target is removed, or whether subject state and scene consistency are maintained when speech content is edited. To address these limitations, AVE-Compass evaluates both edit completion and the preservation of unedited visual and auditory content. It further assesses whether the edited audio and visual signals remain physically and temporally consistent. These capabilities are supported by audio-visually coupled instructions, fine-grained checklists, a dedicated realism rubric, and complementary cross-modal, video, and audio metrics.

%% file: content/3_Benchmark.tex
\section{AVE-Compass: Benchmark and Evaluation}
\label{sec:AVE-Compass}

\begin{figure}[t]
    \centering
    \includegraphics[width=1\linewidth]{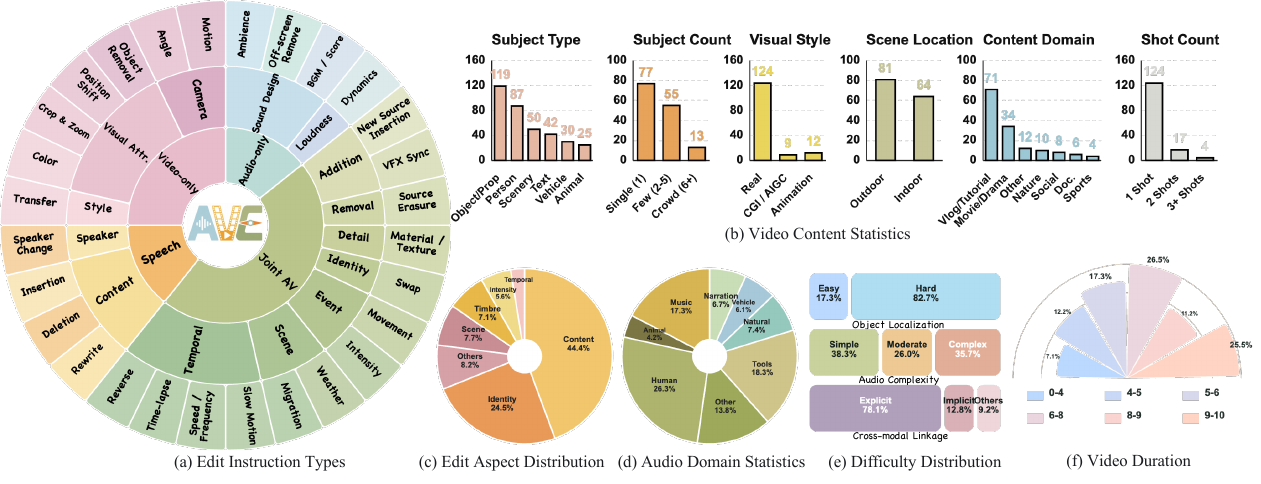}
    \caption{Dataset statistics of the AVE-Compass benchmark.}
    \label{fig:benchmark_stat}
\end{figure}

Fig.~\ref{fig:benchmark_stat} summarizes the statistics of AVE-Compass. Our taxonomy covers a broad range of audio-visual editing operations across joint audio-visual, speech, video-only, and audio-only branches (Fig.~\ref{fig:benchmark_stat}(a)). The source videos span diverse content domains and subject types. They also vary in scene location, shot count, and visual style (Fig.~\ref{fig:benchmark_stat}(b)). The edit-aspect and audio-domain distributions show that the instructions cover content, identity, scene, timbre, temporal, and acoustic editing targets (Fig.~\ref{fig:benchmark_stat}(c,d)). In addition, the difficulty and duration distributions provide varied evaluation conditions (Fig.~\ref{fig:benchmark_stat}(e,f)). Together, these statistics support a diagnostic benchmark for free-form audio-visual editing.

\subsection{Data Construction}
\label{sec:data-construction}

\paragraph{Source Video Collection.}
To construct a diverse source-video pool, we collect candidate samples from existing video datasets (OmniVideoBench~\citep{li2025omnivideobench} and UltraVideo~\citep{xue2025ultravideo}), in-the-wild web videos, and AIGC-generated videos from contemporary systems. We then remove videos with invalid audio tracks or audio-visual misalignment. We also discard samples with ambiguous subjects, ambiguous sound sources, or unclear editable targets. For the retained samples, we annotate the visual content and shot structure, together with subject information, visual style, and audio sources. After filtering and annotation, AVE-Compass contains \textbf{145} source videos.

\paragraph{Edit Instruction Generation.}
To ensure broad task coverage, AVE-Compass organizes these instructions into \textbf{28} fine-grained operation types across four branches: joint audio-visual, speech, video-only, and audio-only editing (Fig.~\ref{fig:benchmark_stat}(a)). The video-only and audio-only branches test non-target modality preservation, while the joint audio-visual and speech branches emphasize cross-modal synchronization and speech-related editing. Beyond operation types, each instruction is further annotated with difficulty-oriented labels from multiple perspectives. These labels describe the audio edit type, object-localization hardness, audio-source complexity, and cross-modal linkage degree. Together, they support stratified analysis across different forms of editing difficulty. Generator prompts and difficulty definitions are provided in Appendix~\ref{app:bench-prompts}.

Covering realistic editing demands, we generate instructions from structured source-video descriptions through a rigorous human-in-the-loop pipeline. First, modality-specific LLM generators propose candidate instructions based on the visual content, audio information, and candidate edit types. A critic model then filters out instructions that are unnatural, infeasible, or insufficiently grounded in the source video. Human annotators perform a final review, and only candidates that pass both stages are retained. This strict selection process yields \textbf{196} human-verified audio-visual editing instructions.

\paragraph{Checklist Generation.}

For fine-grained diagnostic evaluation, we convert each verified instruction into atomic checklist items. Given the source-video description and the editing instruction, checklist generators produce verifiable Yes/No questions that assess both edit execution and the preservation of non-target content. Human annotators then review every checklist item, remove redundant or weakly grounded checks, and revise unclear questions to ensure that each item is specific, verifiable, and aligned with the intended evidence condition. After this careful refinement, the benchmark contains \textbf{2,688} checklist items for MLLM-based evaluation.

\newcommand{\tagvideo}{\fcolorbox{tagblue!70!black}{tagblue!12}{\scriptsize\textbf{Video}}}
\newcommand{\tagaudio}{\fcolorbox{tagred!70!black}{tagred!12}{\scriptsize\textbf{Audio}}}

\begin{table*}[t]
\centering
\small
\setlength{\tabcolsep}{4.0pt}
\renewcommand{\arraystretch}{1.15}
\caption{
Comparison of AVE-Compass with existing editing benchmarks.
\textbf{Target Modality} denotes the evaluated streams.
\textbf{Task Labels} summarize task coverage with normalized labels, including object, scene, motion, camera, VFX, audio event, and speech.
\textbf{\# Metrics} and \textbf{\# Task Categories} count evaluation dimensions and editing operation types, respectively.
\textbf{Cross-Modal Evaluation} indicates whether the benchmark evaluates audio--video coupled editing, ranging from object-level correspondence to global-scene dependencies with diagnostic checks.
\textbf{Multi-Shot} and \textbf{Speech Edit} indicate support for multi-shot source clips and speech-related editing.
\textbf{Difficulty Analysis} denotes stratified evaluation over factors such as target localization, audio-source complexity, and cross-modal linkage.
AVE-Compass is the only benchmark that jointly supports all these capabilities.
}
\label{tab:bench_comparison}
\resizebox{\textwidth}{!}{%
\begin{tabular}{@{}c c c c c c c c c@{}}
\toprule
\textbf{Benchmark}
& \textbf{\begin{tabular}[c]{@{}l@{}}Target\\Modality\end{tabular}}
& \textbf{\begin{tabular}[c]{@{}l@{}}Task\\Labels\end{tabular}}
& \textbf{\begin{tabular}[c]{@{}c@{}}\#\\Metrics\end{tabular}}
& \textbf{\begin{tabular}[c]{@{}c@{}}\# Task\\Categories\end{tabular}}
& \textbf{\begin{tabular}[c]{@{}l@{}}Cross-Modal\\Evaluation\end{tabular}}
& \textbf{\begin{tabular}[c]{@{}c@{}}Multi-\\Shot\end{tabular}}
& \textbf{\begin{tabular}[c]{@{}c@{}}Speech\\Edit\end{tabular}}
& \textbf{\begin{tabular}[c]{@{}l@{}}Difficulty\\Analysis\end{tabular}} \\
\midrule

\rowcolor{groupblue}
\multicolumn{9}{c}{\textit{\textbf{Video-Only Editing Benchmarks}}} \\
\midrule

VEBench~\cite{sun2025ve}
& \tagvideo
& Object, Scene
& 3
& N/A
& \textcolor{crossred}{\ding{55}}
& \textcolor{crossred}{\ding{55}}
& \textcolor{crossred}{\ding{55}}
& \textcolor{crossred}{\ding{55}} \\

\rowcolor{rowcream}
IVEBench~\cite{chen2025ivebench}
& \tagvideo
& Object, Scene
& 12
& 35
& \textcolor{crossred}{\ding{55}}
& \textcolor{crossred}{\ding{55}}
& \textcolor{crossred}{\ding{55}}
& \textcolor{crossred}{\ding{55}} \\

FiVE~\cite{li2025five}
& \tagvideo
& Object
& 15
& 6
& \textcolor{crossred}{\ding{55}}
& \textcolor{crossred}{\ding{55}}
& \textcolor{crossred}{\ding{55}}
& \textcolor{crossred}{\ding{55}} \\

\rowcolor{rowcream}
UniVBench~\cite{wei2026univbench}
& \tagvideo
& Scene
& 21
& 6
& \textcolor{crossred}{\ding{55}}
& \textcolor{checkgreen}{\ding{51}}
& \textcolor{crossred}{\ding{55}}
& \textcolor{crossred}{\ding{55}} \\

CoVEBench~\cite{wu2026covebench}
& \tagvideo
& Object, Scene
& 11
& 19
& \textcolor{crossred}{\ding{55}}
& \textcolor{crossred}{\ding{55}}
& \textcolor{crossred}{\ding{55}}
& \textcolor{crossred}{\ding{55}} \\

\midrule
\rowcolor{groupgreen}
\multicolumn{9}{c}{\textit{\textbf{Audio-Visual Editing Benchmarks}}} \\
\midrule

SAVEBench~\cite{xu2025schrodinger}
& \tagvideo\;\tagaudio
& Object
& 12
& 1
& Object-level
& \textcolor{crossred}{\ding{55}}
& \textcolor{crossred}{\ding{55}}
& \textcolor{crossred}{\ding{55}} \\

\rowcolor{rowcream}
AVED-Bench~\cite{lin2026zero}
& \tagvideo\;\tagaudio
& Object
& 5
& 3
& Object-level
& \textcolor{crossred}{\ding{55}}
& \textcolor{crossred}{\ding{55}}
& \textcolor{crossred}{\ding{55}} \\

AVI-Edit~\cite{zheng2025audio}
& \tagvideo\;\tagaudio
& Audio Event
& 7
& 3
& Object-level
& \textcolor{crossred}{\ding{55}}
& \textcolor{crossred}{\ding{55}}
& \textcolor{crossred}{\ding{55}} \\

\midrule
\rowcolor{highlight}
\textbf{AVE-Compass (Ours)}
& \tagvideo\;\tagaudio
& \textbf{\begin{tabular}[c]{@{}c@{}}Object, Scene,\\Audio Event, Speech\end{tabular}}
& \textbf{19}
& \textbf{28}
& \textbf{Global Scene}
& \textcolor{checkgreen}{\ding{51}}
& \textcolor{checkgreen}{\ding{51}}
& \textcolor{checkgreen}{\ding{51}} \\
\bottomrule
\end{tabular}%
}
\end{table*}

As summarized in Table~\ref{tab:bench_comparison}, prior benchmarks either evaluate visual or audio modalities in isolation, or cover only basic object-level audio-visual operations. AVE-Compass addresses these limitations by requiring cross-modal causality across diverse edit types. It further supports multi-shot source clips, speech-targeted editing, and three-axis difficulty stratification.

\subsection{Evaluation Methodology}
\label{sec:eval-methodology}

As summarized in Tables~\ref{tab:main_subjective} and~\ref{tab:main_objective}, AVE-Compass evaluates edited clips with two complementary components. The MLLM-as-Judge evaluation focuses on semantic edit correctness, source preservation, and perceptual plausibility. The automated metrics provide reproducible measurements of cross-modal synchronization and visual quality. They also cover temporal consistency, audio quality, and speech quality. We report the two components separately so that instruction-level behavior and low-level audio-visual quality are not conflated.

\subsubsection{MLLM-as-Judge Evaluation}
\label{sec:mllm-judge}

We report four dimensions: \textbf{Editing Intent (EI)}, \textbf{Instruction Following (IF)}, \textbf{Fidelity Preserving (FP)}, and \textbf{Realism (REAL)}. As shown in Table~\ref{tab:main_subjective}, each dimension is reported with overall, video, and audio scores.

\paragraph{Editing Intent (EI).} EI rewards an output only when the requested edit is executed and non-target content is preserved. As a result, non-response, incorrect response, and destructive editing all receive low EI scores, making it the primary indicator of complete edit execution in Table~\ref{tab:main_subjective}.

\paragraph{Instruction Following (IF).} We evaluate IF using atomic Yes/No questions derived from each editing instruction. These questions assess whether the requested modification is correctly applied to the intended target and whether the resulting content matches the semantics specified by the instruction. When the instruction includes more detailed constraints, the checklist further examines whether the required attributes, edit extent, spatial or temporal coverage, and removal conditions are satisfied. For visual edits, the questions focus on whether the intended object, region, action, style, or scene is modified at the correct location and to the expected degree. For audio edits, they examine whether the requested changes to sound or speech are accurately realized, including their content, acoustic properties, and temporal placement. For joint audio-visual edits, IF additionally verifies that the changes in both modalities are jointly completed and remain consistent with the coupled relationship expressed in the instruction.

\paragraph{Fidelity Preserving (FP).} We evaluate FP using checklist questions that compare the source and edited videos from two perspectives. First, preservation questions examine whether non-edited visual content remains consistent with the source and whether retained speech, music, ambience, and event sounds are preserved. Second, unintended-content questions assess whether the edited result introduces content that should not appear, such as new objects, subtitles, watermarks, events, narration, background music, or independent sounds. The evaluator ignores trivial differences and direct consequences of the requested edit, but penalizes noticeable changes to retained content and the introduction of clearly unrequested content.

\paragraph{Realism (REAL).} We evaluate the overall naturalness and coherence of each edited clip using a separate MLLM rubric. Unlike the checklists used to assess edit execution, REAL focuses on whether the resulting video appears plausible and natural on its own. On the video side, it evaluates whether the edited elements are visually plausible, physically consistent, and naturally integrated into the surrounding scene. On the audio side, it examines signal quality and whether the acoustic and material properties of the sound are consistent with the visual content and scene context. Checklist generation details and the judging protocol are in Appendix~\ref{app:bench-subjective}.

\subsubsection{Automated Evaluation}
\label{sec:automated-metrics}

Complementing the subjective evaluation, we compute seven automated metrics that provide reproducible scalar measurements. These metrics are organized into \mbox{Cross-Modal}, Video, and Audio groups. All metrics are normalized so that larger values indicate better performance.

\paragraph{Cross-Modal Metrics.} We assess temporal alignment between the edited audio and visual streams using two metrics. For speech-bearing edits, \textbf{Lip Sync} measures whether the edited speech remains synchronized with visible mouth motion using SyncNet confidence. For edits in which sound events must correspond temporally to visual events, \textbf{AV Sync (AVS)} uses Synchformer to estimate the offset between the two modalities and converts it into a normalized score under a fixed tolerance window.

\paragraph{Video Metrics.} We evaluate the edited video in terms of visual quality, subject stability, and temporal coherence. \textbf{Video Aesthetic (V-AES)} averages aesthetic scores over sampled frames to measure the overall visual appeal of the clip. \textbf{Subject Consistency (SC)} computes the cosine similarity between DINOv2 frame features to determine whether the main subject remains stable throughout the edited video. \textbf{Motion Smoothness (MSM)} evaluates whether motion evolves smoothly over time based on frame-interpolation consistency between adjacent frames.

\paragraph{Audio Metrics.} We evaluate both the overall perceptual quality of the edited audio and the quality of its speech content. \textbf{Audio Aesthetic (A-AES)} measures the naturalness and production quality of the complete audio track. For speech-bearing edits, \textbf{Speech Quality (SF)} further evaluates speech intelligibility and naturalness. It predicts a MOS score for the edited speech signal and normalizes the result to $[0,1]$.

The per-category metric activation table and detailed formulas are provided in Appendix~\ref{app:bench-objective}.

%% file: content/4_Agent.tex
\section{Audio-Visual Editing Agent}
\label{sec:agent}

\begin{figure}[t]
    \centering
    \includegraphics[width=0.95\linewidth]{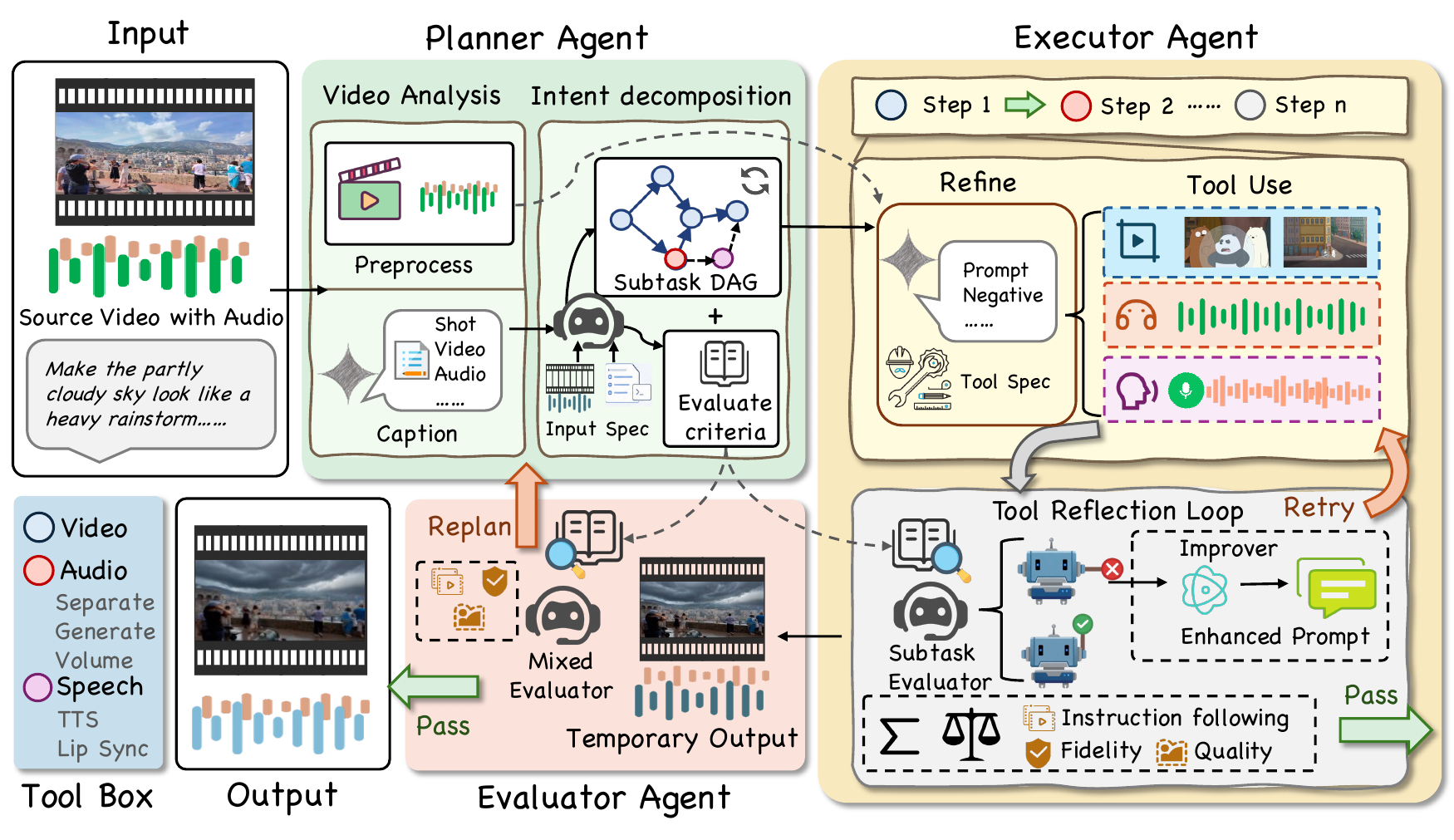}
    \caption{\textbf{Overview of AVE-Agent.} Given a source video with audio and an edit instruction, the planner agent first preprocesses the input audio-visual clip and analyzes it through structured captioning. Conditioned on this analysis, it decomposes the instruction into a dependency-aware subtask DAG with per-step evaluation criteria. The executor agent executes the subtasks in order. For each subtask, it refines the intent into tool prompts and tool specifications, routes the request to the video, audio, or speech branch, and optimizes the result through a reflection loop guided by the planner-provided criteria. Finally, the mixed evaluator agent consumes the temporary assembled clip and planner criteria, scores instruction following, Fidelity Preserving, and quality, and outputs a control signal for final pass, remixing, subtask regeneration, or replanning.}
    \label{fig:agent_arch}
\end{figure}

We build \textbf{AVE-Agent}, a modular pipeline that compiles a free-form natural-language instruction into an edited audio-visual clip~\citep{yao2022react,shen2023hugginggpt,schick2023toolformer}. As shown in Fig.~\ref{fig:agent_arch}, the pipeline runs three agents. The planner agent handles video analysis and intent decomposition, the executor agent performs subtask execution through a self-check reflection loop, and the mixed evaluator agent assesses the assembled clip before replanning. Implementation details, tool choices, and retry budgets are deferred to Appendix~\ref{app:agent}.

\paragraph{Planner agent.}
\label{sec:agent-planner}

The planner agent converts an abstract user instruction into an executable and verifiable editing plan with explicit dependencies. It first builds a structured profile of the input clip, extracting the scene, subjects, actions, and audio context needed for subsequent editing. Conditioned on this profile, a Gemini-based planner identifies not only the explicit requirements in the instruction but also their unstated cross-modal consequences. For example, modifying a visible action may also require updating its associated sound. The planner then decomposes these requirements into fine-grained, modality-specific subtasks and connects them with dependency edges that encode both execution order and audio-visual relations, forming the subtask DAG shown in Fig.~\ref{fig:agent_arch}. Each subtask is paired with actionable evaluation criteria, so downstream execution does not rely on an ambiguous interpretation of the original instruction. Finally, a plan validator checks whether the DAG is complete, executable, and consistent across modalities. Missing dependencies or structural conflicts trigger replanning before any editing tool is invoked.

\paragraph{Executor agent with self-check reflection loop.}
\label{sec:agent-execute}

The executor agent grounds the planned subtasks in concrete editing operations while performing local error correction at every step. Each subtask is first routed to the video, audio, or speech branch according to its type and the priority of available tools. Because a high-level editing intent is rarely suitable as a direct tool input, the executor compiles it into tool-specific guidance, such as a visual editing prompt, a separation target for SAM, an MMAudio request with automatically derived negative constraints, or a speech-editing instruction. After each invocation, a subtask evaluator checks the output against the criteria produced by the planner rather than accepting the first result~\citep{madaan2023self}. When the output fails, an improver uses the evaluator feedback to rewrite the previous guidance and target the identified error in the next attempt, avoiding blind retries from scratch. The loop retains the highest-scoring candidate throughout execution, ensuring that a later unstable attempt cannot overwrite a better result.

\paragraph{Mixed evaluator agent.}
\label{sec:agent-mixeval}

Passing every subtask-level check does not guarantee that the assembled clip is globally coherent, because new conflicts may emerge only after independently edited components are combined. For example, an individually plausible sound effect may obscure dialogue, or a generated audio track may no longer match the modified scene. The mixed evaluator therefore inspects the complete clip for instruction following, fidelity preservation, and overall quality, targeting cross-step failures that local evaluators cannot observe. It then selects the least costly action sufficient to address the detected issue. A mixing problem triggers only a remix, a faulty audio result leads to regeneration of the corresponding subtask, and errors involving plan structure or cross-modal dependencies are sent back to the planner for full replanning with the evaluator feedback as additional context. This minimal-action policy avoids unnecessary recomputation, while a global highest-scoring-output rule prevents regressions across the full retry budget.

%% file: content/5_Experiments.tex
\section{Experiments}
\label{sec:results}

\subsection{Main Results}
\label{sec:results-main}

\begin{table*}[t]
\centering
\caption{\textbf{MLLM-as-Judge results} on AVE-Compass. We report four dimensions---Editing Intent, Instruction Following, Fidelity Preserving, and Realism---each as Overall / Video / Audio scores on a $0$--$100$ scale. Models are ranked by the Overall Editing Intent score, which serves as the primary measure of complete edit execution by jointly accounting for instruction following and fidelity preservation. All metrics are higher-is-better. Best in \textbf{bold}. $^{*}$Gemini misses 16 speech edits due to content moderation.}
\label{tab:main_subjective}
{\setlength{\tabcolsep}{1.6pt}
\resizebox{\textwidth}{!}{
\begin{tabular}{l ccc ccc ccc ccc}
\toprule
\multirow{2}{*}{\textbf{Model}} &
\multicolumn{3}{c}{\textbf{Editing Intent}} &
\multicolumn{3}{c}{\textbf{Instruction Following}} &
\multicolumn{3}{c}{\textbf{Fidelity Preserving}} &
\multicolumn{3}{c}{\textbf{Realism}} \\
\cmidrule(lr){2-4} \cmidrule(lr){5-7} \cmidrule(lr){8-10} \cmidrule(lr){11-13}
 & Overall & Video & Audio & Overall & Video & Audio & Overall & Video & Audio & Overall & Video & Audio \\
\midrule
AVE-Agent (Wan) & \textbf{59.8}$^{+17.4}$ & \textbf{66.7}$^{+6.6}$ & \textbf{50.2}$^{+25.4}$ & \textbf{77.3}$^{+8.0}$ & \textbf{84.8}$^{+6.5}$ & \textbf{69.4}$^{+9.1}$ & 77.6$^{+13.4}$ & 80.1$^{+1.6}$ & 74.8$^{+26.7}$ & 62.1$^{+1.7}$ & 45.3$^{+1.8}$ & 78.9$^{+1.5}$ \\
Wan2.7        & 42.4 & 60.1 & 24.8 & 69.3 & 78.3 & 60.3 & 64.2 & 78.5 & 48.1 & 60.4 & 43.5 & 77.4 \\
\midrule
HappyHorse    & 41.3 & 56.7 & 18.8 & 66.9 & 75.5 & 54.5 & 63.9 & 74.5 & 53.3 & 63.0 & 49.9 & 76.0 \\
Gemini-Omni$^{*}$ & 38.0 & 56.1 & 10.0 & 44.9 & 74.0 & 10.8 & \textbf{84.9} & 76.8 & \textbf{96.1} & 66.9 & 49.2 & 84.5 \\
Seedance      & 26.6 & 36.1 & 13.5 & 37.4 & 50.5 & 24.4 & 81.7 & \textbf{83.0} & 80.4 & \textbf{69.2} & \textbf{54.0} & 84.4 \\
LTX2          & 15.2 & 10.7 & 26.4 & 70.1 & 72.8 & 66.2 & 30.6 & 24.5 & 42.3 & 64.1 & 42.0 & \textbf{86.2} \\
\bottomrule
\end{tabular}
}
}
\vspace{-1mm}
\end{table*}
\begin{table*}[t]
\centering
\caption{\textbf{Automated metric results} on AVE-Compass, grouped into Cross-Modal, Video, and Audio metrics. All metrics are higher-is-better. Best in \textbf{bold}. $^{*}$Gemini misses 16 speech edits due to content moderation. $^{\dagger}$Speech Quality and Lip Sync are computed only on speech-category edits.}
\label{tab:main_objective}
{\setlength{\tabcolsep}{5.6pt}
\resizebox{\textwidth}{!}{
\begin{tabular}{l cc ccc cc}
\toprule
\multirow{2}{*}{\textbf{Model}} &
\multicolumn{2}{c}{\textbf{Cross-Modal}} &
\multicolumn{3}{c}{\textbf{Video}} &
\multicolumn{2}{c}{\textbf{Audio}} \\
\cmidrule(lr){2-3} \cmidrule(lr){4-6} \cmidrule(lr){7-8}
& \makecell{Lip \\ Sync$^{\dagger}$} & \makecell{AV \\ Sync}
& \makecell{Video \\ Aesthetic} & \makecell{Subject \\ Consistency} & \makecell{Motion \\ Smoothness}
& \makecell{Audio \\ Aesthetic} & \makecell{Speech \\ Quality$^{\dagger}$} \\
\midrule
AVE-Agent (Wan) & \textbf{0.622}$^{+0.091}$ & \textbf{0.766}$^{+0.073}$ & 0.452$^{+0.001}$ & 0.972$^{+0.003}$ & 0.987$^{+0.001}$ & 0.614$^{+0.005}$ & 0.368$^{-0.060}$ \\
Wan2.7        & 0.531 & 0.693 & 0.451 & 0.969 & 0.986 & 0.609 & 0.428 \\
\midrule
HappyHorse    & 0.620 & 0.695 & 0.439 & \textbf{0.975} & \textbf{0.989} & \textbf{0.650} & \textbf{0.726} \\
Gemini-Omni$^{*}$ & -- & 0.701 & 0.434 & 0.974 & 0.988 & 0.627 & -- \\
Seedance      & 0.431 & 0.718 & 0.430 & 0.971 & 0.987 & 0.629 & 0.388 \\
LTX2          & 0.618 & 0.758 & \textbf{0.468} & 0.968 & 0.986 & 0.641 & 0.522 \\
\bottomrule
\end{tabular}
}
}
\vspace{-2mm}
\end{table*}

We evaluate six audio-visual editing systems on AVE-Compass, including five baselines---Wan2.7~\citep{wang2025wan}, HappyHorse~\citep{happyhorse2026}, Gemini-Omni~\citep{google2026geminiomni}, Seedance~\citep{gao2025seedance}, and LTX2.3-Retake (LTX2)~\citep{hacohen2024ltx}---as well as AVE-Agent (Wan). Tables~\ref{tab:main_subjective} and~\ref{tab:main_objective} present the MLLM-as-Judge and automated metric results, respectively. Despite its compact size, the benchmark contains 196 instructions that collectively cover all operation categories and difficulty levels defined in our taxonomy, providing sufficient breadth for diagnostic evaluation. Conducting substantially larger-scale evaluations would also be expensive for routine use, as discussed in Appendix~\ref{app:limitations}.

The results reveal three main trends.
(1) \textbf{AVE-Agent is the strongest edit executor.} It achieves the best performance on Editing Intent and Instruction Following, with particularly large gains in audio-side instruction following and audio-visual synchronization. Beyond these leading results, AVE-Agent also outperforms the baselines on most of the remaining metrics, demonstrating the effectiveness of the agent-based workflow.
(2) \textbf{Current models struggle to balance Instruction Following and Fidelity Preserving.} Models that perform relatively well on Instruction Following often do so at the expense of preservation. For instance, LTX2 follows many instructions but largely regenerates the input video, resulting in poor Fidelity Preserving. In contrast, models with high Fidelity Preserving may fail to carry out the requested edit. Gemini-Omni and Seedance, for example, sometimes return the original video or audio, which yields high preservation scores while leaving the instruction unsatisfied. We therefore define Editing Intent as IF$\times$FP to capture complete edit execution. As shown in Table~\ref{tab:main_subjective}, existing models still achieve relatively low scores on this metric.
(3) \textbf{Audio-visual realism remains limited.} Although some models achieve high Video Aesthetic or Audio Aesthetic scores, their Realism scores remain noticeably lower than their automated quality scores. Automated metrics primarily measure the technical quality of the generated content, while Realism evaluates the logical validity and audio-visual naturalness of the edited result. This discrepancy suggests that current models continue to struggle with physical-world understanding in the context of editing.

\subsection{Further Analysis}
\label{sec:results-finegrained}

\paragraph{Edit Response and Gated Fidelity Preserving.}
High Fidelity Preserving or Realism scores do not necessarily indicate successful editing, as they may arise when a model leaves the input unchanged. To separate preservation quality from edit non-response, Table~\ref{tab:gated_fidelity} reports modality-specific response rates and recomputes Fidelity Preserving and Realism only over cases in which the target modality is actually modified. A non-response is still counted as an Instruction Following failure, but is excluded from the gated preservation and Realism averages. This analysis reveals two different forms of inaction. Gemini rarely modifies the audio track, artificially increasing its ungated Audio Fidelity Preserving score, whereas Seedance more frequently leaves the visual stream unchanged. AVE-Agent, by contrast, exhibits the smallest overall decrease after gating, indicating that its preservation and Realism performance is less attributable to unchanged outputs.

\paragraph{Robustness under Difficulty.}
We further examine robustness across source-video duration, object-localization hardness, audio-source complexity, and cross-modal linkage degree in Figure~\ref{fig:robustness}, with complete results provided in Appendix~\ref{subsec:difficulty_full}. Longer source videos reduce Editing Intent for most baselines, whereas AVE-Agent remains comparatively stable. For each of the three instruction-level difficulty dimensions, we report the metric most directly affected by that factor. Increasing object-localization hardness weakens fine-grained target grounding, while greater audio-source complexity makes it more difficult to preserve non-target content in acoustically crowded scenes. Higher cross-modal linkage degree further exposes failures to infer and execute implicit audio edits associated with visual changes. Gemini-Omni and Seedance obtain relatively high Audio Fidelity Preserving scores under complex audio conditions, but these values are partly inflated by audio non-response, consistent with the gated analysis above. Overall, the results show that current systems remain sensitive to distinct sources of editing difficulty, while AVE-Agent maintains more stable performance across these conditions.

\begin{table*}[t]
\centering
\caption{\textbf{Response-gated Fidelity Preserving and Realism} by modality. Response rate measures whether the target modality changes. The ``all'' columns average over all evaluated cases, while the ``gated'' columns recompute Fidelity Preserving and Realism only on cases where the corresponding target modality responds. All metrics are higher-is-better. \textbf{Bold} marks the lowest, i.e., worst, response rates.}
\label{tab:gated_fidelity}
\resizebox{\textwidth}{!}{
\begin{tabular}{l cc cc cc cc cc}
\toprule
\multirow{2}{*}{\textbf{Model}} &
\multicolumn{2}{c}{\textbf{Response Rate}} &
\multicolumn{2}{c}{\textbf{Fidelity Preserving (all)}} &
\multicolumn{2}{c}{\textbf{Fidelity Preserving (gated)}} &
\multicolumn{2}{c}{\textbf{Realism (all)}} &
\multicolumn{2}{c}{\textbf{Realism (gated)}} \\
\cmidrule(lr){2-3}\cmidrule(lr){4-5}\cmidrule(lr){6-7}\cmidrule(lr){8-9}\cmidrule(lr){10-11}
 & Video & Audio & Video & Audio & Video & Audio & Video & Audio & Video & Audio \\
\midrule
AVE-Agent (Wan) & 92.7 & 94.4 & 80.1 & 74.8 & 77.9 & 70.9 & 45.3 & 78.9 & 43.0 & 76.6 \\
Wan2.7      & 90.6 & 95.6 & 78.5 & 48.1 & 76.3 & 44.8 & 43.5 & 77.4 & 40.4 & 75.1 \\
\midrule
HappyHorse  & 95.8 & 74.4 & 74.5 & 53.3 & 73.1 & 31.7 & 49.9 & 76.0 & 48.8 & 73.2 \\
Gemini-Omni & 89.7 & \textbf{22.0} & 76.8 & 96.1 & 74.2 & 80.6 & 49.2 & 84.5 & 46.8 & 81.2 \\
Seedance    & \textbf{69.1} & 47.8 & 83.0 & 80.4 & 74.9 & 60.1 & 54.0 & 84.4 & 48.5 & 82.8 \\
LTX2        & 90.1 & 91.1 & 24.5 & 42.3 & 16.8 & 39.0 & 42.0 & 86.2 & 36.7 & 84.4 \\
\bottomrule
\end{tabular}
}
\end{table*}

\begin{figure*}[t]
\centering
\includegraphics[width=\textwidth]{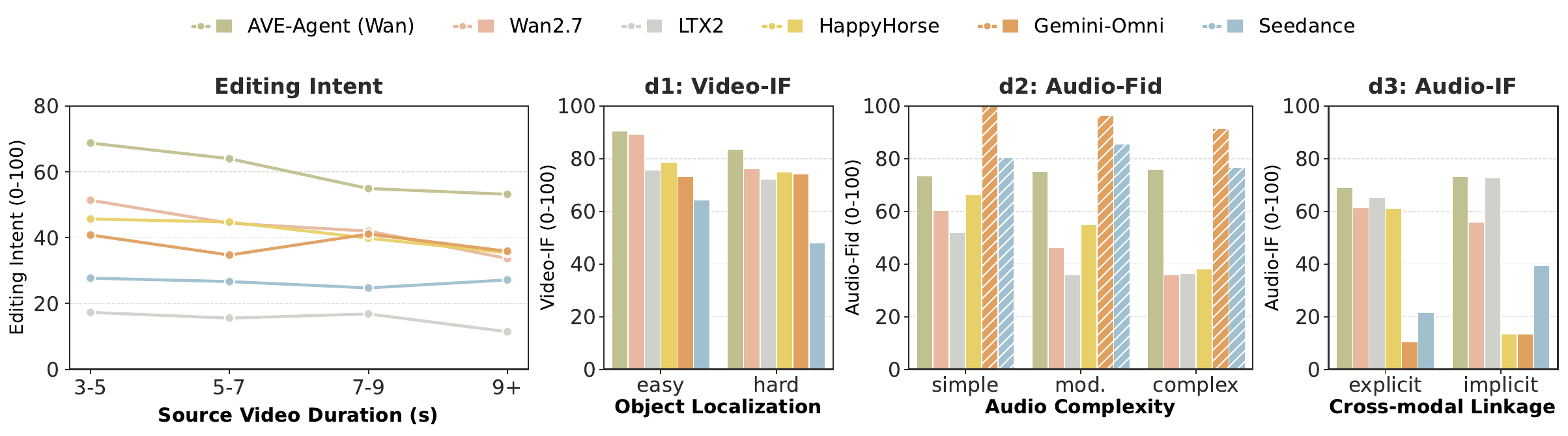}
\caption{\textbf{Robustness under difficulty.} Left: Editing Intent across source-video duration. Right: object-localization hardness, audio-source complexity, and cross-modal linkage degree evaluated with modality-matched metrics. Hatched audio-source complexity bars mark Audio Fidelity Preserving inflated by audio non-response. Scores are on a $0$--$100$ scale.}
\label{fig:robustness}
\end{figure*}

\paragraph{Analysis of Metric Redundancy and Orthogonality.}

\begin{figure*}[t]
    \centering
    \includegraphics[width=\textwidth]{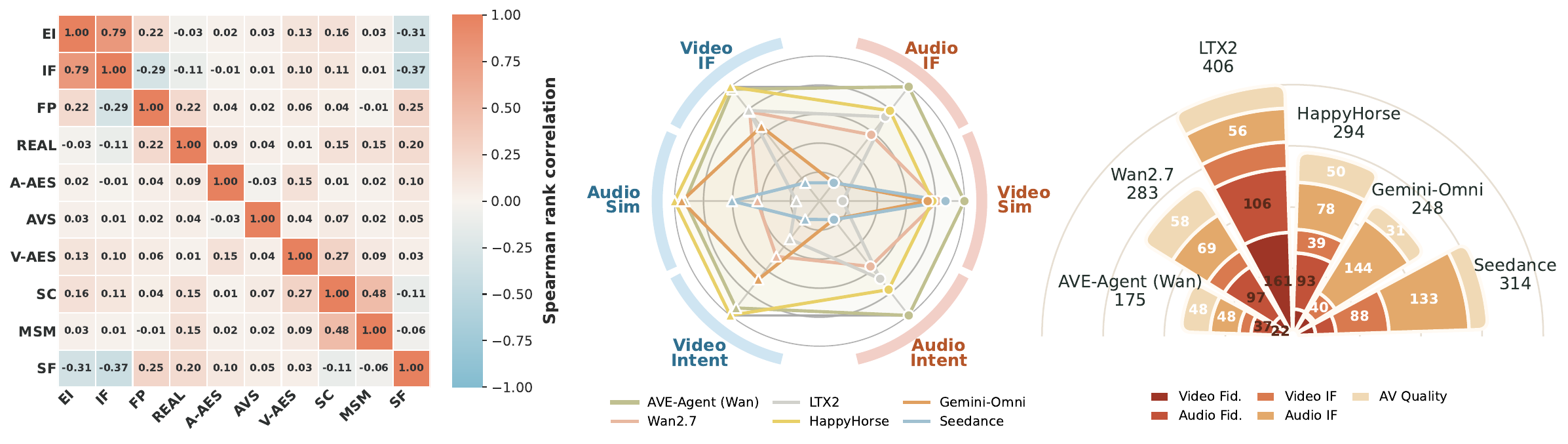}
    \caption{\textbf{Metric orthogonality, single-modality editing, and error analysis.} Left: Spearman correlation between subjective and objective metrics. Middle: video-only and audio-only performance, with axes normalized per metric. Right: per-model error counts across five failure categories.}
    \label{fig:single_modality}
\end{figure*}

The left panel of Figure~\ref{fig:single_modality} presents the Spearman correlations between objective and subjective metrics. Correlations across the two groups are generally weak, suggesting that they capture largely distinct yet complementary aspects of editing quality and are thus approximately orthogonal. In particular, low-level objective metrics may fail to detect realism defects identified by the subjective rubric, such as deformed hands, disappearing objects, or edited regions that do not blend naturally into the surrounding scene. By contrast, several objective metrics exhibit strong correlations with one another, revealing partial redundancy among low-level signal-based measures. Together, these observations show that neither metric group alone provides a complete assessment and motivate our multi-dimensional evaluation framework rather than reliance on a single aggregate score.

\paragraph{Metric Validity Evaluation.}

We evaluate metric validity from two complementary perspectives: repeated-run stability and consistency with human judgments. To assess repeated-run stability, we execute the complete scoring protocol five times on a representative subset covering all six evaluated models. Across these runs, the automated metric scores vary by less than 0.01, while the MLLM-as-Judge scores vary by less than $1\%$. This limited variation indicates that the reported model comparisons are robust to evaluator stochasticity. We further assess human consistency by comparing the MLLM-as-Judge results with human annotations on sampled questions, with the calculation procedure detailed in Appendix~\ref{subsec:human_consistency_calculation}. Reported in Table~\ref{tab:human_llm_agreement}, the average agreement remains close to $90\%$ across Editing Intent, Instruction Following, Fidelity Preserving, and Realism, providing further evidence for the reliability of the automatic evaluation.

\begin{table*}[t]
\centering
\caption{\textbf{Human-LLM agreement} of MLLM-as-Judge metrics. Agreement measures consistency between human annotations and automatic MLLM judgments, averaged over evaluated models. Values are percentages.}
\label{tab:human_llm_agreement}
\resizebox{\textwidth}{!}{
\begin{tabular}{lcccccccccccc}
\toprule
\multirow{2}{*}{\textbf{Metric}} &
\multicolumn{3}{c}{\textbf{Editing Intent}} &
\multicolumn{3}{c}{\textbf{Instruction Following}} &
\multicolumn{3}{c}{\textbf{Fidelity Preserving}} &
\multicolumn{3}{c}{\textbf{Realism}} \\
\cmidrule(lr){2-4}\cmidrule(lr){5-7}\cmidrule(lr){8-10}\cmidrule(lr){11-13}
& Overall & Video & Audio & Overall & Video & Audio & Overall & Video & Audio & Overall & Video & Audio \\
\midrule
\textbf{Agreement} & 89.9 & 90.0 & 89.9 & 91.3 & 92.9 & 89.7 & 88.8 & 87.9 & 90.1 & 91.0 & 93.0 & 89.0 \\
\bottomrule
\end{tabular}
}
\end{table*}

\paragraph{Single-Modality Editors on AVE-Compass.}

We further evaluate single-modality editing, where only the target stream should be modified and the other modality should remain unchanged. The middle panel of Figure~\ref{fig:single_modality} presents the main results, with the complete numbers reported in Appendix~\ref{subsec:single_modality_full}. To measure preservation of the non-target modality more directly, we additionally report automated Audio Similarity and Video Similarity scores. For video-only edits, AVE-Agent and HappyHorse both perform strongly by modifying the visual content while retaining the source audio. For audio-only edits, AVE-Agent achieves the best overall performance, successfully following the audio instruction without altering the video. By contrast, some non-responsive baselines obtain deceptively high preservation scores simply because they leave the target stream unchanged.

\paragraph{Error Analysis and Case Study}
Figure~\ref{fig:intro} illustrates the central difficulty faced by current models: executing the requested edit while preserving non-target content across both modalities. The qualitative examples in Appendix Figure~\ref{fig:editing_comparison} further reveal distinct model-specific failure patterns. Some systems, such as LTX2, extensively re-synthesize the source video or audio and thereby damage content that should remain unchanged. Others, such as Gemini-Omni, may leave the target modality largely unedited or introduce implausible visual artifacts and unsupported content. These cases show that reliable audio-visual editing requires coordinated edit execution and content preservation in both streams.

\textbf{Quantitative error breakdown.} To quantify these failure patterns, we assign each edit to five concrete error categories. The right panel of Figure~\ref{fig:single_modality} summarizes their frequencies, while Appendix Table~\ref{tab:error_analysis} provides the complete breakdown. AVE-Agent produces the fewest errors overall, without being dominated by any single failure type. By contrast, models prone to non-response fail primarily on Instruction Following, whereas highly regenerative models fail mainly on Fidelity Preserving. Gemini's low number of Audio Fidelity Preserving errors should therefore not be interpreted as strong preservation performance, as it largely results from its tendency not to edit the audio stream.

\subsection{Ablation Studies}
\label{sec:results-ablation}

\begin{figure*}[t]
    \centering
    \includegraphics[width=\textwidth]{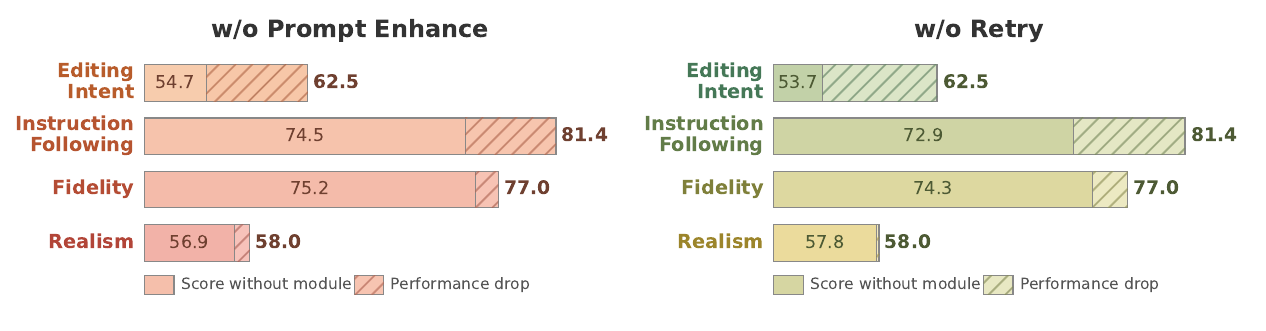}
    \caption{Ablation of planner-side prompt enhancement and evaluator-driven retry refinement. Solid bars show the score without the module; hatched segments show the drop from the full AVE-Agent.}
    \label{fig:ablation_evaluation}
\end{figure*}

Figure~\ref{fig:ablation_evaluation} separately ablates two agent components: planner-side prompt enhancement and evaluator-driven retry refinement. Removing prompt enhancement decreases Editing Intent by $7.76$ and Instruction Following by $6.91$, highlighting the importance of translating user intent into executable tool prompts before invoking specialized editors. The accompanying decline in Realism further indicates that prompt enhancement also improves the quality of the edited outputs. Removing retry has an even larger impact on edit execution, lowering Editing Intent by $8.75$ and Instruction Following by $8.51$, as failed or incomplete edits can no longer be revised using evaluator feedback. Fidelity Preserving also drops, suggesting that retry refinement contributes to maintaining non-target content during execution. Overall, prompt enhancement improves the quality of the initial tool invocation, whereas retry refinement provides the corrective mechanism required for reliable editing. Appendix Tables~\ref{tab:agent_backbone_subjective} and~\ref{tab:agent_backbone_objective} further report the agent-backbone ablation results.

\label{sec:results-discussion}

%% file: content/6_Conclusion.tex
\section{Conclusion}

In this paper, we introduce AVE-Compass, a comprehensive benchmark for coordinated audio-visual editing. By moving beyond the visual-only focus of existing benchmarks, AVE-Compass systematically evaluates Instruction Following, Fidelity Preserving, Realism, and Editing Intent. Its evaluation framework combines checklist-based MLLM judging, a dedicated realism rubric, and automated metrics for cross-modal, video, and audio quality. We further propose AVE-Agent, a modular framework that addresses complex cross-modal editing through structured task planning and iterative self-reflection. Our experiments uncover the characteristic failure modes of current models and show that AVE-Agent achieves the strongest edit-execution performance while maintaining competitive preservation and perceptual quality. We hope AVE-Compass will help advance the field from isolated pixel-level modifications toward physically and perceptually coherent multimodal editing.

%% file: content/Appendix_Benchmark.tex
\section{Benchmark Construction Details}

This appendix reports the objective and subjective evaluation formulas used by AVE-Compass. The latest prompt templates for source-video captioning, checklist generation, checklist evaluation, and realism evaluation are collected separately in Appendix~\ref{app:prompt-templates}.

\subsection{Objective Metric Formulas}
\label{app:bench-objective}

This appendix details the objective branch of AVE-Compass. Let the source video be
$x=(v,a)$, where $v$ and $a$ denote the visual and audio streams, and let the edited
output be $\hat{x}=(\hat{v},\hat{a})$ under instruction $p$. Each sample is assigned
one instruction category
$c\in\{\text{audio-only},\text{video-only},\text{joint},\text{speech}\}$.
The category determines which objective metrics are active. Inactive metrics are
excluded rather than assigned zero, so that a video-only edit is not penalized by
audio-target metrics and an audio-only edit is not penalized by visual-target metrics.
Unless otherwise noted, each score is oriented so that a larger value is better.
Predictor scores with bounded native ranges are mapped to $[0,1]$. Similarity scores
based on correlations or cosine similarities are also reported as larger-is-better
scores, with affine mapping applied when the native score lies in $[-1,1]$.

Table~\ref{tab:AVE-Compass_objective_dimensions} lists the seven reference-free
objective scores reported in the main experiment table. Additional source-target
similarity scores for single-modality preservation are described after the table.

\begin{table*}[t]
\centering
\small
\setlength{\tabcolsep}{4.0pt}
\renewcommand{\arraystretch}{1.15}
\caption{Reference-free objective score routing in AVE-Compass. Each score is activated only for the instruction categories where it provides relevant evidence.}
\label{tab:AVE-Compass_objective_dimensions}
\resizebox{\textwidth}{!}{%
\begin{tabular}{p{0.24\linewidth}p{0.36\linewidth}p{0.34\linewidth}}
\toprule
\textbf{Metric Group} & \textbf{Score Name} & \textbf{Active Categories} \\
\midrule
\multirow{2}{*}{\textbf{Cross-Modal}}
& lip\_sync & speech \\
& av\_sync & audio-only, joint, speech \\
\midrule
\multirow{3}{*}{\textbf{Video}}
& video\_aesthetic & video-only, joint, speech \\
& subject\_consistency & video-only, joint, speech \\
& motion\_smoothness & video-only, joint, speech \\
\midrule
\multirow{2}{*}{\textbf{Audio}}
& audio\_aesthetic & audio-only, joint, speech \\
& speech\_quality & speech \\
\bottomrule
\end{tabular}%
}
\end{table*}

For single-modality diagnostics, AVE-Compass additionally reports preservation
similarities on the non-target stream: \texttt{videoclip\_video\_similarity} for
audio-only edits and \texttt{audio\_similarity} for video-only edits.

\paragraph{Audio aesthetic.}
\texttt{audio\_aesthetic} measures the perceptual quality of the edited audio track.
We extract the audio from $\hat{x}$ with ffmpeg as 16\,kHz, mono, PCM 16-bit WAV and
evaluate it with AudioBox Aesthetics via the T2AV-Compass
\texttt{run\_audiobox\_batch.py} script. The model predicts Perceptual Quality (PQ),
Content Understanding (CU), Content Enjoyment (CE), and Production Complexity (PC).
Following the implementation, we average PQ and CU and divide by 10:
\begin{equation}
S_{\mathrm{AA}}=\frac{\mathrm{PQ}+\mathrm{CU}}{20}.
\end{equation}
The resulting score lies in $[0,1]$, where higher values indicate cleaner, more
natural, and semantically coherent audio.

\paragraph{Speech quality.}
\texttt{speech\_quality} is enabled only for \texttt{speech} edits. It measures the
quality and intelligibility of the edited speech signal with NISQA, a non-intrusive
speech quality assessment model. The edited audio is extracted from $\hat{x}$ and
evaluated with \texttt{run\_predict.py --mode predict\_dir}. NISQA returns a predicted
mean opinion score (MOS) together with diagnostic scores for noise, distortion,
coloration, and loudness. We use MOS as the primary score and map the native range
$[1,5]$ to $[0,1]$:
\begin{equation}
S_{\mathrm{speech}}=\operatorname{clip}\left(\frac{\mathrm{MOS}-1}{4},0,1\right).
\end{equation}
Higher values indicate more natural, intelligible, and artifact-free speech.

\paragraph{Lip synchronisation.}
\texttt{lip\_sync} is enabled for \texttt{speech} edits, where speech generation or
speech modification should remain aligned with visible mouth motion. The evaluator
uses SyncNet through \texttt{batch\_lipsync.py}. For an edited video $\hat{x}$,
SyncNet extracts visual lip features and speech features, searches over temporal
shifts, and returns an audio--visual frame offset, a minimum feature distance, and a
confidence score. The implementation uses the confidence value
$\kappa=\mathrm{median\_dist}-\mathrm{min\_dist}$ and normalizes it by a configurable
threshold $\gamma$:
\begin{equation}
S_{\mathrm{lip}}=\operatorname{clip}\left(\frac{\kappa}{\gamma},0,1\right),
\qquad \gamma=2.0.
\end{equation}
Larger values indicate stronger agreement between the edited speech and the visible
mouth movements.

\paragraph{Audio--visual synchronisation.}
\texttt{av\_sync} evaluates whether the edited audio and visual streams are
temporally aligned. The implementation uses Synchformer, either through a preloaded
model path or through the T2AV-Compass \texttt{batch\_av\_sync.py} fallback. The model
predicts the temporal offset $\Delta t$ in seconds between audio and video. We convert
the offset to a normalized score with tolerance $\tau=2.0$\,s:
\begin{equation}
S_{\mathrm{sync}}=\max\left(0,\ 1-\frac{|\Delta t|}{\tau}\right),\qquad \tau=2.0.
\end{equation}
An exactly aligned sample receives 1.0; the score decreases linearly with absolute
offset and becomes 0 once the predicted offset reaches 2 seconds.

\paragraph{Video aesthetic.}
\texttt{video\_aesthetic} measures the visual aesthetic quality of the edited video.
The edited video is evaluated by Aesthetic Predictor v2.5, a CLIP-based image
aesthetic model, through \texttt{batch\_video\_aesthetic.py}. The evaluator uniformly
samples $T=10$ frames from $\hat{v}$, scores each frame, averages the raw frame-level
scores, and divides by 10:
\begin{equation}
S_{\mathrm{VA}}=\frac{1}{10T}\sum_{i=1}^{T}q_i,\qquad T=10,
\end{equation}
where $q_i$ is the raw aesthetic score of the $i$-th sampled frame. The score is in
$[0,1]$, with larger values indicating stronger visual appeal and fewer aesthetic
defects.

\paragraph{Subject consistency.}
\texttt{subject\_consistency} measures whether the main subject appearance remains
stable across frames in the edited video. It is computed on $\hat{v}$ only and does
not use the source video. The evaluator uniformly samples at most 50 frames, resizes
each frame to $224\times224$, normalizes it with ImageNet statistics, and extracts
DINOv2 features. For each adjacent frame pair, it computes the cosine similarity and
then maps the mean adjacent-frame cosine from $[-1,1]$ to $[0,1]$:
\begin{equation}
\bar{c}_{\mathrm{adj}}=\frac{1}{T-1}\sum_{i=2}^{T}\cos(f_i,f_{i-1}),
\qquad
S_{\mathrm{subj}}=\operatorname{clip}_{[0,1]}\left(\frac{\bar{c}_{\mathrm{adj}}+1}{2}\right).
\end{equation}
The implementation also reports the mean pairwise cosine across all sampled frame
pairs as a diagnostic field, but the official score is the adjacent-frame score above.
Higher values indicate that subjects, objects, and their visual identity remain more
temporally consistent after editing.

\paragraph{Motion smoothness.}
\texttt{motion\_smoothness} detects unnatural motion, flicker, or temporal
discontinuities in the edited video. The evaluator extracts frames from $\hat{v}$ and
uses the even-indexed frames $(0,2,4,\ldots)$ as inputs to AMT-G, a video frame
interpolation model. AMT-G predicts the missing middle frames at temporal midpoint
$0.5$. These predicted odd frames are compared with the actual odd-indexed frames
$(1,3,5,\ldots)$ by pixel-wise mean absolute difference:
\begin{equation}
d_i=\operatorname{mean}\left(|I^{\mathrm{interp}}_i-I^{\mathrm{real}}_i|\right),
\qquad
d=\frac{1}{N}\sum_{i=1}^{N}d_i.
\end{equation}
The raw difference $d$ lies on the 8-bit pixel scale and is converted to a
larger-is-better score:
\begin{equation}
S_{\mathrm{motion}}=\frac{255-d}{255}.
\end{equation}
The implementation adapts resolution and padding according to available GPU memory
before AMT-G inference. Larger scores indicate smoother and more predictable motion.

\paragraph{VideoCLIP video similarity.}
\texttt{videoclip\_video\_similarity} is used for audio-only edits, where
the visual stream should remain unchanged while the audio is edited. The evaluator
uniformly samples 8 frames from both $v$ and $\hat{v}$, resizes them to
$224\times224$, normalizes them with ImageNet statistics, and feeds each 8-frame tube
into the VideoCLIP-XL visual encoder. The source and edited video embeddings are
L2-normalized and compared by cosine similarity, then mapped to $[0,1]$:
\begin{equation}
c_{\mathrm{vclip}}=\cos\left(\phi_v(v),\phi_v(\hat{v})\right),
\qquad
S_{\mathrm{vclip}}=\operatorname{clip}_{[0,1]}\left(\frac{c_{\mathrm{vclip}}+1}{2}\right).
\end{equation}
If OpenCV cannot seek reliably, for example for some AV1-encoded videos, the
implementation falls back to ffmpeg-based frame extraction. The raw cosine is kept in
the metric details for inspection, while the reported score is the normalized value;
higher values indicate stronger preservation of the original visual content.

\paragraph{Spectral audio similarity.}
\texttt{audio\_similarity} is used for video-only edits, where the audio track should
remain stable while the visual stream is edited. We extract audio from the source and
edited videos with ffmpeg, convert both tracks to 16\,kHz mono PCM WAV, and compute
STFT magnitude spectrograms $M(a)$ and $M(\hat{a})$. The default window size is
1024 samples with a half-window hop. To tolerate small timing shifts introduced by
editing or encoding, the evaluator searches over spectrogram-frame lags within a
configurable window, $3$\,s by default, while requiring sufficient overlap between
the two clips.

For a candidate lag $\delta$, both overlapping spectrogram regions are flattened and
z-score normalized before Pearson correlation is computed. The best lag
$\delta^\ast$ maximizes this spectral correlation:
\begin{equation}
\delta^\ast=\arg\max_{\delta}
\rho\left(M(a),M(\hat{a})_{\delta}\right),
\qquad
S_{\mathrm{audio}}=\operatorname{clip}\left(\frac{1+\rho\left(M(a),M(\hat{a})_{\delta^\ast}\right)}{2},0,1\right).
\end{equation}
If both extracted tracks are near-silent, the implementation treats them as preserved
and returns 1.0. Higher values indicate that the original audio content and acoustic
structure are preserved, while lower values indicate unintended drift, deletion,
replacement, or severe distortion.

\subsection{Subjective Evaluation Formulas}
\label{app:bench-subjective}

This appendix documents the latest checklist-based MLLM-as-Judge protocol introduced in Sec.~\ref{sec:mllm-judge}. The checklist no longer uses a fixed 15-question layout or embeds Realism as one of the checklist dimensions. Instead, each checklist contains atomic Yes/No questions for \textbf{Instruction Following} and \textbf{Fidelity Preserving}, plus separate \textbf{Edit Response} questions that measure whether the model attempted an edit in the relevant modality. Realism is evaluated by the standalone realism prompt reported in Appendix~\ref{app:prompt-realism-evaluation}.

\paragraph{Atomic checklist generation.}
For every instruction, a branch-specific generator produces a concise checklist grounded in the source caption and edit prompt. Video-only, audio-only, joint audio-video, and speech edits use separate generators, but all follow the same schema. Each item carries a dimension tag, a subdimension tag, and a modality tag. The scored checklist dimensions are \emph{Instruction Following}, which checks whether the requested edit semantics, target, attributes, coverage, and removal/replacement conditions are satisfied, and \emph{Fidelity Preserving}, which checks preservation of non-edited content and control of unrequested hallucinations. The number of questions is adaptive rather than fixed: the generator keeps only the most informative atomic checks for the specific edit.

\paragraph{Edit Response questions.}
The checklist includes additional response-rate questions with dimension \emph{Edit Response} and subdimension \texttt{edit\_response}. These questions are descriptive, not correctness-oriented. They ask only whether the target contains a visible or audible change compared with the source. Video-only edits include a video response question, audio-only edits include an audio response question, and joint audio-video edits include both video and audio response questions. Speech edits always include an audio response question and add a video response question only when the speech edit is expected to affect mouth motion, subtitles, speaker identity, or other visual content. Edit Response answers are reported as response-rate statistics and are excluded from the Instruction Following and Fidelity Preserving scores.

\paragraph{Absolute Affirmative Paradigm.}
Every scored question is phrased so that a successful edit answers ``Yes''. Instead of a negatively oriented probe such as ``Is the removed object still visible?'', the checklist asks the success-state form ``Is the requested region free of the removed object?''. The MLLM is instructed to answer ``Yes'' only when the evidence clearly satisfies the exact condition, and to answer ``No'' for absent, ambiguous, partial, weak, or approximate matches. This eliminates polarity ambiguity in removal and replacement tasks and disallows trivially true compound questions.

\paragraph{Modality-split (Tripartite Call) judging.}
To curb cross-modal bias, the MLLM is invoked separately on each modality. Audio-tagged questions receive only source and target audio; video-tagged questions receive only source and target video; general questions receive integrated audio--video streams so that synchronisation and coupled events can still be judged jointly. Each split call returns the original question ID, dimension, subdimension, modality tag, observation, binary answer, and short justification; per-question outputs are merged by ID.

\paragraph{Item representation.}
Each checklist item is represented as $q_i=(\mathrm{id}_i, d_i, s_i, t_i, r_i)$, where $\mathrm{id}_i$ is the question ID, $d_i\in\{\mathrm{ER},\mathrm{IF},\mathrm{FP}\}$ is the dimension tag for Edit Response, Instruction Following, or Fidelity Preserving, $s_i$ is the subdimension tag, $t_i\in\{\mathrm{audio},\mathrm{video},\mathrm{general}\}$ is the modality tag, and $r_i$ is the natural-language question.

\paragraph{Human-in-the-loop deduplication.}
After generation, human annotators run a conservative deduplication pass: items are merged only when they would receive the same Yes/No answer under the same evidence; annotators retain the more specific question and renumber the rest, never adding, rewriting, or answering items. Hence the deduplicated checklist may contain fewer items than the raw generated checklist,
\begin{equation}
\mathcal{Q}^{\prime}=\mathrm{HumanDedup}(\mathcal{Q}),\qquad |\mathcal{Q}^{\prime}|\leq |\mathcal{Q}|.
\end{equation}

\paragraph{Binarised answers and per-modality scores.}
Following the Absolute Affirmative Paradigm, each MLLM answer is binarised as
\begin{equation}
y_i=
\begin{cases}
1, & \text{if the answer to } q_i \text{ is Yes},\\
0, & \text{if the answer to } q_i \text{ is No}.
\end{cases}
\end{equation}
For a dimension $d\in\{\mathrm{IF},\mathrm{FP},\mathrm{ER}\}$ and a modality set $M\subseteq\{\mathrm{audio},\mathrm{video},\mathrm{general}\}$, let
\begin{equation}
\mathcal{I}_{d,M}=\{i\mid q_i\in\mathcal{Q}^{\prime},\ d_i=d,\ t_i\in M\}.
\end{equation}
The per-case score is the Yes-rate over the corresponding items,
\begin{equation}
S_{d,M}=
\frac{1}{|\mathcal{I}_{d,M}|}\sum_{i\in\mathcal{I}_{d,M}} y_i,
\qquad
\mathcal{I}_{d,M}\neq\emptyset .
\end{equation}
AVE-Compass reports video-side scores with $M=\{\mathrm{video}\}$, audio-side scores with $M=\{\mathrm{audio}\}$, and overall scores with $M=\{\mathrm{video},\mathrm{audio},\mathrm{general}\}$. Empty modality--dimension cells are omitted rather than imputed.

\paragraph{Editing Intent score.}
Instruction Following and Fidelity Preserving are complementary: a model must both carry out the requested edit and preserve non-target content. Therefore the intent score is computed by a per-case product rather than by averaging the two dimensions independently. For case $c$ and modality set $M$,
\begin{equation}
S_{\mathrm{Intent},M}^{(c)}
=S_{\mathrm{IF},M}^{(c)}\cdot S_{\mathrm{FP},M}^{(c)}.
\end{equation}
Audio intent and video intent are obtained by using $M=\{\mathrm{audio}\}$ and $M=\{\mathrm{video}\}$, respectively. Dataset-level results are macro-averaged over cases where both factors are defined,
\begin{equation}
\bar{S}_{\mathrm{Intent},M}
=\frac{1}{|\mathcal{C}_M|}
\sum_{c\in\mathcal{C}_M}
S_{\mathrm{IF},M}^{(c)}S_{\mathrm{FP},M}^{(c)}.
\end{equation}
The same macro-averaging rule is used for Instruction Following, Fidelity Preserving, and Edit Response scores.

%% file: content/Appendix_Agent.tex
\section{Implementation Details of AVE-Agent}
\label{app:agent}

This appendix documents the implementation of the three modules introduced in Sec.~\ref{sec:agent}. We first describe the Video Analyzer that produces the input spec and the typed schemas used by the planner. We then detail the tool registry, the per-tool self-check loop, and the Mixed Evaluator's actionable signals and retry budgets.

\subsection{Video Analyzer}
\label{app:agent-analyzer}

Preprocessing probes metadata and separates the audio and visual tracks. It then applies HDR$\to$SDR tone-mapping, normalises resolution to a $720$--$2160$\,px width band with square pixel aspect, and extracts $1$--$3$ keyframes. The captioner runs Gemini~2.5~Flash on the full clip and emits a structured caption with a visual paragraph, an audio paragraph, and a shot list, where shots are camera setups rather than jump cuts.

\subsection{Planner Schema}
\label{app:agent-planner}

\textbf{Phase~A intent fields.} A Gemini~2.5~Flash planner emits an ordered list of typed intents, each with an \texttt{action} (e.g.\ \texttt{replace\_object}, \texttt{audio\_replace\_sfx}, \texttt{speech\_tts}), a \texttt{shot\_index} or global flag, a \texttt{depends\_on} list realising the visual$\to$audio causal order, and a per-intent audio block listing \texttt{existing\_sounds}, \texttt{deleted\_sound}, and \texttt{new\_sound}. A \texttt{PlanValidator} enforces audio$\leftrightarrow$visual consistency (e.g.\ a removal must name its target inside \texttt{existing\_sounds}; a vocal action must emit both a visual \texttt{motion\_edit} and an \texttt{audio\_*} intent), triggering up to two automatic replans on violation.

\textbf{Phase~B output fields.} Each validated intent is independently translated in parallel into modality-specific prompt fields (\texttt{video\_prompt}, \texttt{sam\_prompt}, \texttt{mmaudio\_prompt}, \texttt{speech\_*}) plus optional per-step eval criteria. Format constraints are strict: $6$--$10$ imperative-verb words for \texttt{video\_prompt}; $\leq 2$ synonyms within $\leq 10$ words for \texttt{sam\_prompt}; sensory description without imperatives for \texttt{mmaudio\_prompt}. MMAudio negative prompts are auto-derived as $\texttt{existing\_sounds} \setminus \{\texttt{deleted\_sound}\}$ rather than written by the LLM.

\subsection{Tool Registry}
\label{app:agent-tools}

Each Subtask is dispatched through a \texttt{ToolRegistry} that selects tools by priority. The \emph{video} branch provides two selectable fal.ai backends: Wan~2.7 is the default V2V editor, while Seedance~2.0 provides reference-to-video regeneration as an alternative~\citep{wang2025wan,gao2025seedance}; shots shorter than the $3$\,s V2V minimum are freeze-padded then re-sliced. The \emph{audio} branch dispatches by edit type into separate / generate / volume sub-steps. It performs source separation with fal.ai SAM~Audio (AudioSep as local fallback)~\citep{shi2025sam,liu2024separate} and visual-conditioned generation with MMAudio~V2 (\texttt{guidance\_scale} $7.0$ for additions, $4.5$ for replacements)~\citep{cheng2025mmaudio}. The \emph{speech} branch decomposes speech edits into atomic \texttt{speech\_tts}, \texttt{speech\_swap}, and per-shot \texttt{speech\_lipsync} steps. These steps use SAM keep-mode reference $\to$ Qwen3 Voice Clone, Qwen3 Voice Design without reference, and Sync~Lipsync~2 respectively~\citep{chung2016out,prajwal2020lip}.

\subsection{Self-Check / Optimization Loop}
\label{app:agent-loop}

Three specialised sub-evaluators back the loop, and each scores against the audio inventory. The \emph{separation} evaluator scores SAM outputs on \texttt{target\_extraction} and \texttt{residual\_fidelity}, with a hard ceiling at $0.4$ when fidelity drops below $0.4$. The \emph{generation} evaluator scores MMAudio outputs on \texttt{content\_present}, \texttt{negative\_violation}, and \texttt{hallucination}; it also returns actionable \texttt{missing}/\texttt{unwanted} fields. The \emph{post-branch} evaluator scores the in-Subtask mix on instruction/sync/fidelity and raises three flags (\texttt{generated\_audio\_too\_quiet}, \texttt{original\_audio\_too\_quiet}, \texttt{generated\_audio\_contamination}) that drive an in-branch volume retry.

Improvers refine rather than reset prompts: the SAM improver preserves the source noun and adds at most $1$--$2$ descriptors; the MMAudio improver only adjusts the terms named in \texttt{missing}/\texttt{unwanted} while accumulating an explicit negative list capped at eight entries. Both improvers retain $\geq 70\%$ of the original prompt and apply hard word caps ($10$ for SAM, $12$ for MMAudio). The Subtask retry budget is up to three attempts.

\subsection{Mixed Evaluator}
\label{app:agent-mix}

The Mixed Evaluator scores the assembled clip item-by-item against the audio inventory. It evaluates instruction following, fidelity, and quality before emitting one of three actionable signals. A \texttt{volume\_adjustment} triggers an ffmpeg-based re-mux only, with no model re-invocation, up to two retries. A \texttt{needs\_regenerate} flag re-runs the last audio Subtask through the same Self-Check / Optimization Loop, up to one retry. When the evaluator surfaces structural problems that no current step can resolve, the pipeline replans by re-entering Intent Decomposition with the evaluator feedback as extra context. Across the full retry budget the pipeline always returns the highest-scoring artefact rather than the most recent one.

%% file: content/Appendix_Experiment.tex
\section{Experiment Details}
\label{app:Experiment}

\subsection{Metric Similarity and Redundancy Analysis}
\label{subsec:metric_redundancy}

To better understand the relationships and potential overlaps among the evaluation metrics, we conduct a metric similarity analysis based on Spearman correlation. The analysis compares the automated metrics with the MLLM-as-Judge dimensions, providing an intuitive view of how different metrics align with one another and with human judgments.

In this analysis, certain objective metrics show meaningful alignment with subjective human preferences, while several objective metrics exhibit exceptionally high correlation coefficients with each other. This strong inter-metric correlation indicates that these metrics capture highly overlapping characteristics when assessing generated outputs. Consequently, it reveals significant \textit{metric redundancy} within the evaluation framework. Employing numerous highly correlated metrics does not necessarily yield additional informational gain regarding model performance. Recognizing and analyzing these redundancies is a crucial step toward streamlining future evaluation protocols. It allows researchers to filter out homogeneous metrics and select a core set that is concise, independent, and representative without sacrificing evaluation comprehensiveness.

The corresponding heatmap is shown in the left panel of Figure~\ref{fig:single_modality} in the main text.

\subsection{Human Consistency Calculation}
\label{subsec:human_consistency_calculation}

We conduct a human consistency study to verify whether the MLLM-as-Judge results are stable with respect to human annotations. For each sampled question, the automatic judgment is counted as \textit{correct} when it is consistent with the human annotation and as \textit{incorrect} otherwise. The human consistency score is then computed as:
\begin{equation}
\mathrm{Consistency}
= \frac{N_{\mathrm{correct}}}
{N_{\mathrm{correct}} + N_{\mathrm{incorrect}}}
\times 100\%,
\end{equation}
where $N_{\mathrm{correct}}$ and $N_{\mathrm{incorrect}}$ denote the numbers of matched and mismatched judgments, respectively. We compute this score for each evaluation dimension and modality, and report the averaged percentages in Table~\ref{tab:human_llm_agreement} of the main text.

\subsection{Difficulty-Stratified Analysis}
\label{subsec:difficulty_full}
Table~\ref{tab:difficulty} gives the full difficulty-stratified breakdown referenced in Sec.~\ref{sec:results-finegrained} (Robustness under Difficulty), each axis paired with its modality-matched metric.

\begin{table*}[t]
\centering
\caption{\textbf{Difficulty-stratified analysis} on AVE-Compass, each axis paired with its modality-matched metric. All values are on a $0$--$100$ scale. Video Realism is normalized from the $1$--$5$ rubric via $(x{-}1)/4\times100$. Baselines degrade with difficulty, while our agent stays notably more robust under audio-source complexity and implicit cross-modal linkage.}
\label{tab:difficulty}
\resizebox{\textwidth}{!}{
\begin{tabular}{l cc cc ccc cc}
\toprule
\multirow{2}{*}{\textbf{Model}} &
\multicolumn{2}{c}{\makecell{\textbf{Object-Localization Hardness}\\\textbf{Video-IF}}} & \multicolumn{2}{c}{\makecell{\textbf{Object-Localization Hardness}\\\textbf{Video Realism}}} &
\multicolumn{3}{c}{\makecell{\textbf{Audio-Source Complexity}\\\textbf{Audio Fidelity Preserving}}} & \multicolumn{2}{c}{\makecell{\textbf{Cross-Modal Linkage}\\\textbf{Audio-IF}}} \\
\cmidrule(lr){2-3} \cmidrule(lr){4-5} \cmidrule(lr){6-8} \cmidrule(lr){9-10}
 & easy & hard & easy & hard & simple & mod. & complex & explicit & implicit \\
\midrule
AVE-Agent (Wan) & 90.5 & 83.7 & 48.5 & 44.7 & 73.6 & 75.1 & 75.9 & 69.0 & 73.3 \\
Wan2.7      & 89.3 & 76.3 & 47.3 & 42.7 & 60.3 & 46.3 & 36.0 & 61.3 & 56.0 \\
\midrule
HappyHorse  & 78.6 & 74.9 & 56.6 & 48.5 & 66.2 & 55.0 & 38.0 & 61.3 & 13.3 \\
Gemini-Omni & 73.1 & 74.2 & 54.4 & 48.1 & 100.0 & 96.4 & 91.4 & 10.5 & 13.3 \\
Seedance    & 64.3 & 48.1 & 59.8 & 52.7 & 80.5 & 85.5 & 76.8 & 21.6 & 39.3 \\
LTX2        & 75.6 & 72.2 & 37.3 & 43.0 & 52.0 & 35.8 & 36.4 & 65.4 & 72.7 \\
\bottomrule
\end{tabular}
}
\end{table*}

\subsection{Single-Modality Editing}
\label{subsec:single_modality_full}
Table~\ref{tab:single_modality} reports the objective single-modality numbers summarized by the radar in the middle panel of Figure~\ref{fig:single_modality}.

\begin{table*}[t]
\centering
\caption{\textbf{Single-modality editing} (objective metrics). \emph{Top}: video-only ($n{=}16$)---visual quality plus Audio Similarity (preservation of the untouched audio). \emph{Bottom}: audio-only ($n{=}5$)---audio quality plus Video Similarity (preservation of the untouched video). Best in \textbf{bold}. $^{\ddagger}$Our agent's Audio Similarity is mildly depressed by a residual $\sim$1-frame muxing offset (frame-level alignment leaves an $\sim$8\,ms residual); sample-aligned, the audio is essentially preserved.}
\label{tab:single_modality}
\begin{minipage}[t]{0.54\textwidth}
\centering
\small
\caption*{\textbf{Video-only editing}}
\resizebox{\linewidth}{!}{%
\begin{tabular}{l cccc}
\toprule
\textbf{Model} & \makecell{Video \\ Aesthetic} & \makecell{Subject \\ Consistency} & \makecell{Motion \\ Smoothness} & \makecell{Audio \\ Similarity} \\
\midrule
AVE-Agent (Wan) & 0.463 & 0.983 & 0.985 & 0.969$^{\ddagger}$ \\
Wan2.7      & 0.467 & 0.983 & 0.986 & 0.759 \\
\midrule
HappyHorse  & 0.454 & \textbf{0.986} & \textbf{0.989} & \textbf{0.999} \\
Gemini-Omni & 0.438 & 0.983 & 0.987 & 0.978 \\
Seedance    & 0.442 & 0.980 & 0.985 & 0.833 \\
LTX2        & \textbf{0.478} & 0.973 & 0.983 & 0.647 \\
\bottomrule
\end{tabular}%
}
\end{minipage}\hfill
\begin{minipage}[t]{0.42\textwidth}
\centering
\small
\caption*{\textbf{Audio-only editing}}
\resizebox{\linewidth}{!}{%
\begin{tabular}{l ccc}
\toprule
\textbf{Model} & \makecell{Audio \\ Aesthetic} & \makecell{AV \\ Sync} & \makecell{Video \\ Similarity} \\
\midrule
AVE-Agent (Wan) & 0.629 & 0.580 & \textbf{0.999} \\
Wan2.7      & 0.608 & 0.520 & 0.983 \\
\midrule
HappyHorse  & 0.594 & \textbf{0.660} & 0.980 \\
Gemini-Omni & 0.581 & 0.340 & 0.977 \\
Seedance    & 0.604 & 0.580 & 0.988 \\
LTX2        & \textbf{0.653} & \textbf{0.660} & 0.924 \\
\bottomrule
\end{tabular}%
}
\end{minipage}
\end{table*}

\subsection{Error Analysis}

To gain deeper insights into the strengths and limitations of our approach, we conduct a comprehensive error analysis across all six models (Table~\ref{tab:error_analysis}; visualized in the right panel of Figure~\ref{fig:single_modality}). We categorize generation defects into five failure categories. These categories cover video/audio Fidelity Preserving (FP), video/audio Instruction Following (IF), and audio-visual quality. An edit is counted as an error in a category when it fails the corresponding checklist questions (per-case yes-rate $<0.5$) or, for audio-visual quality, when its Realism score falls below $3/5$. Each value is therefore an \textbf{error count}---the number of edits failing that category---so lower is better.

Our agent incurs the fewest total errors ($175$, versus $248$--$406$ for the baselines), and the per-category composition reveals each model's characteristic failure mode. The non-response baselines fail predominantly on Instruction Following: Audio-IF accounts for $144$ of Gemini's and $133$ of Seedance's errors, as they frequently return the source unedited. Notably, Gemini's very low Audio Fidelity Preserving error count ($5$) is \emph{not} a strength but a direct consequence of this non-response---it cannot corrupt audio it never edits. The regenerative baselines fail instead on Fidelity Preserving because they re-synthesize rather than preserve the source content. For example, LTX2 has $161$ video-FP and $106$ audio-FP errors, while Wan and HappyHorse have $97$ and $93$ audio-FP errors respectively. Our agent is the only model with no dominant failure category. Its two largest categories, Audio-IF and audio-visual quality, are tied ($48$ each), with the latter consistent with the visual-quality headroom of the Wan backend discussed in the main text.

\begin{table*}[t]
    \centering
    \caption{\textbf{Error analysis across five failure categories} (error counts; lower is better). An edit is counted as an error in a category when it fails the corresponding checklist questions (per-case yes-rate $<0.5$) or, for audio-visual quality, when its Realism is below $3/5$. Best (lowest) per row in \textbf{bold}; Gemini's low Audio Fidelity Preserving count reflects non-response, not preservation quality.}
    \label{tab:error_analysis}
    \small
    \renewcommand{\arraystretch}{1.2}
    \setlength{\tabcolsep}{3pt}
    \begin{tabular}{p{0.30\textwidth} c c c c c c}
        \toprule
        \textbf{Category} & \makecell{\textbf{AVE-Agent}\\\textbf{(Wan)}} & \textbf{Wan2.7} & \makecell{\textbf{Happy}\\\textbf{Horse}} & \makecell{\textbf{Gemini}\\\textbf{Omni}} & \textbf{Seedance} & \textbf{LTX2} \\
        \midrule
        1. Poor Video Fidelity Preserving (FP$\times$video) & \textbf{22} & 27 & 34 & 28 & 28 & 161 \\
        2. Poor Audio Fidelity Preserving (FP$\times$audio) & 37 & 97 & 93 & \textbf{5} & 34 & 106 \\
        3. Poor Video Instruction Following (IF$\times$video) & \textbf{20} & 32 & 39 & 40 & 88 & 44 \\
        4. Poor Audio Instruction Following (IF$\times$audio) & \textbf{48} & 69 & 78 & 144 & 133 & 56 \\
        5. Poor AV Quality (Realism $<3/5$) & 48 & 58 & 50 & \textbf{31} & \textbf{31} & 39 \\
        \midrule
        \textbf{Total} & \textbf{175} & 283 & 294 & 248 & 314 & 406 \\
        \bottomrule
    \end{tabular}
\end{table*}

\subsection{Case Analysis}

\begin{figure*}[htbp]
    \centering
    \includegraphics[width=\textwidth]{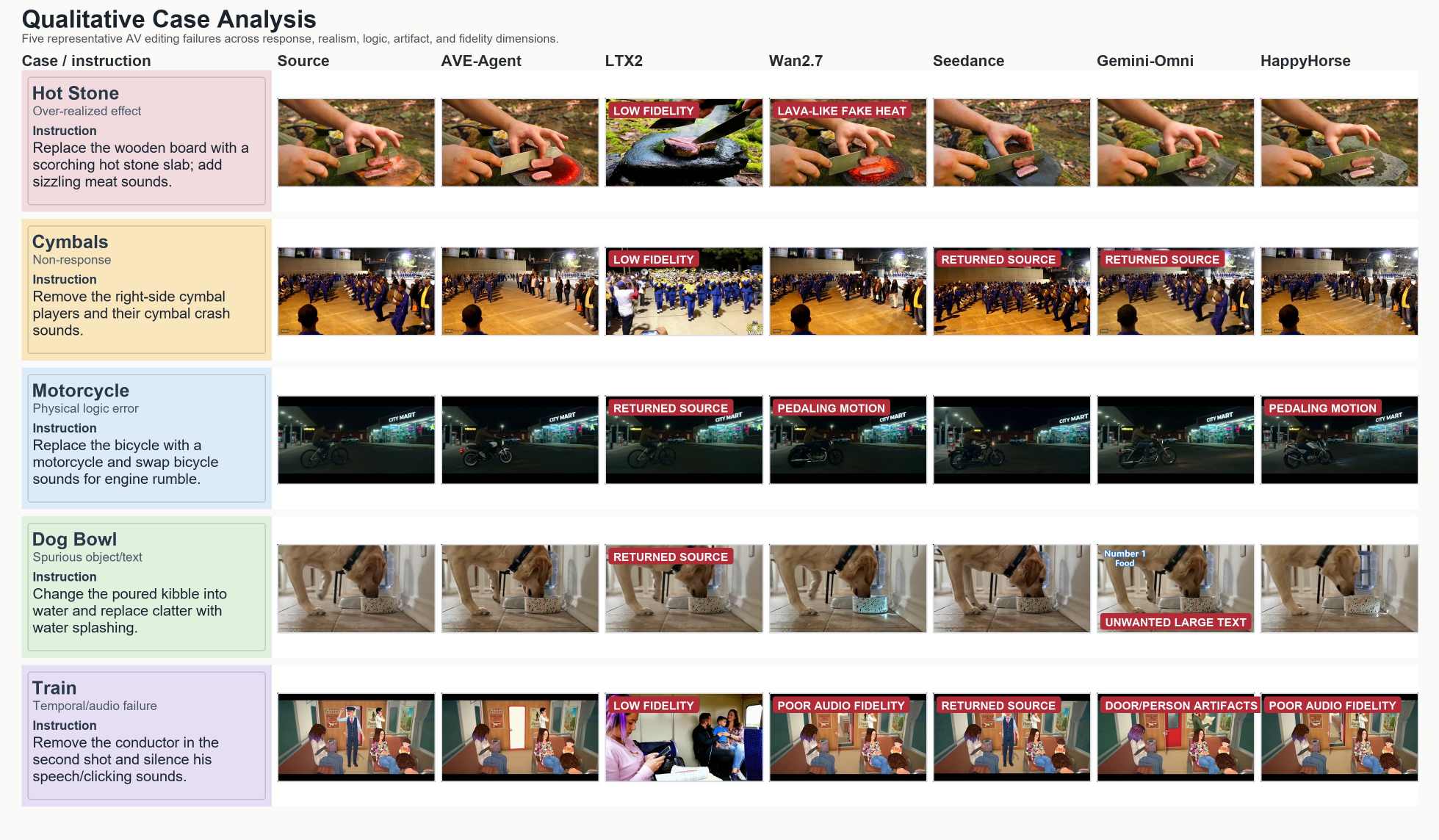}
    \caption{\textbf{Qualitative case analysis.} Each row shows the source frame, edit instruction, and aligned edited frames from AVE-Agent, LTX2, Wan2.7, Seedance, Gemini-Omni, and HappyHorse. Red tags mark the dominant failure observed in each selected output, including low Fidelity Preserving under regeneration, over-realistic visual effects, non-response, physical-logic errors, spurious text/object insertion, temporal inconsistency, and poor Audio Fidelity Preserving.}
    \label{fig:editing_comparison}
\end{figure*}

Figure~\ref{fig:editing_comparison} summarizes the characteristic weaknesses of each baseline with representative cases. \textbf{Wan2.7} often struggles with Audio Fidelity Preserving and can produce visually over-synthetic edits. In \textbf{\path{RmyA2l-3WQU_58.0_66.0_manual.1}}, its conductor-removal result substantially degrades the audio. In \textbf{\path{05f8sG4OhZs_787.0_800.0_manual.1}}, it turns the requested hot stone slab into a lava-like, AI-looking effect. \textbf{LTX2} has two failure modes. A portion of cases are not edited at all, as in \textbf{\path{SR__amDl1c8_10.0_17.0_manual.1}} and \textbf{\path{7ik2wgDkvxc_31.0_38.0_manual.1}}. When it does edit, its retake-style regeneration changes the source video beyond recognition, leading to extremely poor Fidelity Preserving, as shown in the hot-stone, cymbal-removal, and train-conductor cases. \textbf{Seedance} frequently fails by non-response. In \textbf{\path{1vy-ZxTMQf4_377.0_387.0_vggsounder_music_performance.1}}, it leaves the right-side cymbal players and their sounds unchanged. In \textbf{\path{RmyA2l-3WQU_58.0_66.0_manual.1}}, it similarly returns the source instead of removing the conductor. \textbf{Gemini-Omni} is prone to editing hallucinations, producing outcomes that violate common sense or introduce unsupported elements. It adds an unrequested large caption in \textbf{\path{7ik2wgDkvxc_31.0_38.0_manual.1}} and creates temporally inconsistent artifacts such as disappearing/reappearing background characters and an implausibly opening train door in \textbf{\path{RmyA2l-3WQU_58.0_66.0_manual.1}}. \textbf{HappyHorse} also frequently exhibits editing hallucinations. In \textbf{\path{SR__amDl1c8_10.0_17.0_manual.1}}, it replaces the bicycle with a motorcycle but keeps bicycle-like pedaling motion, producing a physically inconsistent result; in the train-conductor case it additionally suffers from poor Audio Fidelity Preserving.

\subsection{Effect of the Agent Backbone}
Tables~\ref{tab:agent_backbone_subjective} and~\ref{tab:agent_backbone_objective} compare the standalone backbones with their agent-augmented variants on the 34-case ablation subset. Superscripts report the difference between each agent and its corresponding base model.
For the Seedance backbone, AVE-Agent (sd2) substantially improves Editing Intent and Instruction Following but reduces Fidelity Preserving and Realism. This is mainly because the standalone Seedance baseline often leaves the input unchanged, which preserves the original video's realism without satisfying the edit. The agent version executes more instructions, but Seedance edits can introduce logical or visual inconsistencies after the requested change, leading to lower preservation and Realism scores.

\begin{table*}[t]
\centering
\caption{\textbf{MLLM-as-Judge agent-backbone ablation} on the 34-case subset. Scores are on a $0$--$100$ scale. All metrics are higher-is-better. Best results within each backbone pair are in \textbf{bold}.}
\label{tab:agent_backbone_subjective}
\resizebox{\textwidth}{!}{
\begin{tabular}{l ccc ccc ccc ccc}
\toprule
\multirow{2}{*}{\textbf{Model}} &
\multicolumn{3}{c}{\textbf{Editing Intent}} &
\multicolumn{3}{c}{\textbf{Instruction Following}} &
\multicolumn{3}{c}{\textbf{Fidelity Preserving}} &
\multicolumn{3}{c}{\textbf{Realism}} \\
\cmidrule(lr){2-4} \cmidrule(lr){5-7} \cmidrule(lr){8-10} \cmidrule(lr){11-13}
 & Overall & Video & Audio & Overall & Video & Audio & Overall & Video & Audio & Overall & Video & Audio \\
\midrule
AVE-Agent (Wan) & \textbf{62.5}$^{+17.1}$ & \textbf{69.4}$^{+10.7}$ & \textbf{57.0}$^{+30.5}$ & \textbf{81.8}$^{+7.7}$ & \textbf{85.3}$^{+2.0}$ & \textbf{79.8}$^{+14.1}$ & \textbf{77.1}$^{+12.2}$ & \textbf{80.9}$^{+8.1}$ & \textbf{73.5}$^{+22.8}$ & \textbf{58.0}$^{+0.1}$ & \textbf{37.7}$^{+1.7}$ & 78.3$^{-1.5}$ \\
Wan & 45.4 & 58.7 & 26.5 & 74.1 & 83.3 & 65.7 & 64.9 & 72.8 & 50.7 & 57.9 & 36.0 & \textbf{79.8} \\
\midrule
AVE-Agent (sd2) & \textbf{49.0}$^{+16.2}$ & \textbf{44.2}$^{+10.5}$ & \textbf{52.0}$^{+25.1}$ & \textbf{70.0}$^{+26.3}$ & \textbf{69.6}$^{+19.1}$ & \textbf{68.5}$^{+31.9}$ & 70.0$^{-12.6}$ & 64.7$^{-17.5}$ & 77.2$^{-6.1}$ & 64.8$^{-8.4}$ & 50.2$^{-10.1}$ & 80.3$^{-5.7}$ \\
Seedance & 32.8 & 33.7 & 26.9 & 43.7 & 50.5 & 36.6 & \textbf{82.6} & \textbf{82.2} & \textbf{83.3} & \textbf{73.2} & \textbf{60.3} & \textbf{86.0} \\
\bottomrule
\end{tabular}
}
\end{table*}

\begin{table*}[t]
\centering
\caption{\textbf{Automated metric agent-backbone ablation} on the 34-case subset. All metrics are higher-is-better. Best results within each backbone pair are in \textbf{bold}. $^{\dagger}$Speech Quality and Lip Sync are computed only on the two speech-category edits.}
\label{tab:agent_backbone_objective}
{\setlength{\tabcolsep}{3.6pt}
\resizebox{\textwidth}{!}{
\begin{tabular}{l cc ccc cc}
\toprule
\multirow{2}{*}{\textbf{Model}} &
\multicolumn{2}{c}{\textbf{Cross-Modal}} &
\multicolumn{3}{c}{\textbf{Video}} &
\multicolumn{2}{c}{\textbf{Audio}} \\
\cmidrule(lr){2-3} \cmidrule(lr){4-6} \cmidrule(lr){7-8}
& \makecell{Lip Sync$^{\dagger}$} & \makecell{AV Sync}
& \makecell{Video Aesthetic} & \makecell{Subject Consistency} & \makecell{Motion Smoothness}
& \makecell{Audio Aesthetic} & \makecell{Speech Quality$^{\dagger}$} \\
\midrule
AVE-Agent (Wan) & \textbf{0.858}$^{+0.060}$ & \textbf{0.821}$^{+0.109}$ & \textbf{0.452}$^{+0.000}$ & \textbf{0.971}$^{+0.004}$ & \textbf{0.986}$^{+0.000}$ & \textbf{0.627}$^{+0.031}$ & 0.402$^{-0.144}$ \\
Wan & 0.798 & 0.712 & \textbf{0.452} & 0.967 & \textbf{0.986} & 0.596 & \textbf{0.546} \\
\midrule
AVE-Agent (sd2) & \textbf{1.000}$^{+0.671}$ & \textbf{0.691}$^{+0.059}$ & 0.419$^{-0.015}$ & \textbf{0.976}$^{+0.007}$ & \textbf{0.989}$^{+0.001}$ & 0.608$^{-0.026}$ & \textbf{0.506}$^{+0.010}$ \\
Seedance & 0.329 & 0.632 & \textbf{0.434} & 0.969 & 0.988 & \textbf{0.634} & 0.496 \\
\bottomrule
\end{tabular}
}
}
\end{table*}

%% file: content/Appendix_BroaderImpact.tex
\section{Broader Impact}
\label{app:broader-impact}

AVE-Compass and AVE-Agent lower the barrier to free-form audio-visual content creation. This can broaden access for education, independent creators, and accessibility-oriented use cases such as re-dubbing and content localisation. They also share the dual-use risks of generative media: better cross-modal coherence can be misused for deepfakes, disinformation, or non-consensual voice cloning, particularly in the speech-targeted branch. To mitigate this risk, our benchmark uses only publicly licensed and AIGC clips, and AVE-Agent ships without identity-cloning tools. We will release the benchmark and code under licences that prohibit generating non-consensual or deceptive media. Downstream users are encouraged to combine the system with provenance signals such as watermarking and content credentials (e.g., C2PA) and with detectors trained on the failure modes surfaced by AVE-Compass.

%% file: content/Appendix_Limitations.tex
\section{Limitations}
\label{app:limitations}

The scale of AVE-Compass reflects a deliberate trade-off against evaluation cost: scoring each clip requires running closed-source video and audio editing APIs as well as MLLM-based checklist judging, all of which are expensive on a per-instance basis. Within this budget, we have intentionally maximised coverage rather than raw count. The benchmark spans 28 fine-grained edit subcategories well beyond subject-level operations and curates source videos across diverse subjects, visual styles, multi-shot structures, and speech / non-speech conditions. This coverage helps AVE-Compass approximate the real-world distribution of audio-visual editing demands. AVE-Agent additionally depends on third-party tools whose behaviour may shift over time, which can affect long-term reproducibility.

As a rough cost reference, consider evaluating one video-editing model with the official BytePlus Seedance 2.0 API plan. The official ModelArk pricing lists Seedance 2.0 at about \$0.76 per 5-second 720P video, or roughly \$0.15 per generated second.\footnote{\url{https://docs.byteplus.com/en/docs/ModelArk/1544106}} With 196 instructions and an average clip duration of roughly 7.4 seconds, one full generation pass costs about \$220 in generation fees ($196 \times 7.4 \times 0.15$). The MLLM judging cost is smaller but still non-negligible. Gemini 3.1 Pro Preview is priced at \$2 per million input tokens and \$12 per million output tokens for prompts below 200k tokens.\footnote{\url{https://ai.google.dev/gemini-api/docs/gemini-3}} Using Google's multimodal tokenization rule of 263 tokens per second for video and 32 tokens per second for audio,\footnote{\url{https://ai.google.dev/gemini-api/docs/tokens}} judging source--edited pairs for checklist and realism evaluation gives an order-of-magnitude budget of a few million input tokens plus a few hundred thousand output tokens, or roughly \$10--15 for one model. A single-model AVE-Compass pass is therefore on the order of \$230--235 in API fees before failed generations, retries, storage, and human verification are counted. This makes the current scale a practical compromise: it covers the full taxonomy for diagnostic comparison, while a substantially larger benchmark would become difficult to run as a routine evaluation.

%% file: content/Appendix_Prompt.tex
\section{Prompt Templates}
\label{app:prompt-templates}

This appendix collects the latest prompt templates used by AVE-Compass. It includes prompts for source-video captioning, checklist generation, checklist evaluation, and realism evaluation, together with short descriptions of their roles in the benchmark pipeline.

\subsection{Source-Video Captioning Prompt}
\label{app:prompt-captioning}

The captioning stage runs a multimodal model on the original video with audio and requests a structured JSON description for downstream prompt generation and evaluation.

\subsubsection{Captioning System Message}
\begin{lstlisting}[style=promptstyle]
SYSTEM_MSG = "Return ONLY valid JSON. No markdown, no commentary."
\end{lstlisting}

\subsubsection{Captioning User Prompt}
\begin{lstlisting}[style=promptstyle]
Return ONLY valid JSON. No extra text.

ROLE & TASK
You are a professional, detail-oriented video analyst. Your job is to describe a video clip WITH AUDIO as comprehensively as possible.
The video may contain one or multiple camera shots / scene cuts. If multiple shots exist, you MUST identify and describe shot transitions within your description.
Radical objectivity: Only write undeniable, verifiable audiovisual facts. If uncertain, write "unclear" or "unknown". Do NOT guess. Do NOT fabricate.

OUTPUT JSON schema (MUST follow exactly):
{
  "caption": "A highly detailed, multi-sentence description of the VISUAL content of the entire clip. If the video contains multiple shots, describe shot-by-shot in chronological order using transition phrases ('The clip opens with...', 'Cut to...', 'The shot transitions to...', 'In the next shot...'). If single-shot, describe the progression of action chronologically. Focus ONLY on what is SEEN --- do NOT describe audio here.",
  "audio_caption": "A highly detailed, multi-sentence description of the AUDIO content of the entire clip as a coherent paragraph. Describe chronologically everything that is HEARD: all speech (who speaks --- describe speaker by appearance if visible, tone, emotion, pace, volume, language, whether dialogue/monologue/narration/voiceover --- but do NOT transcribe words), sound effects tied to on-screen events (footsteps, impacts, doors, explosions, water, etc.), ambient/environmental sounds (crowd noise, wind, traffic, room tone, etc.), and background music (genre, mood, instrumentation, tempo, intensity, how it evolves --- rises, fades, shifts key/mood). Describe how audio elements layer, interact, and change across the clip.",
  "speech_or_non_speech": "speech|non_speech",
  "shot_count": 0,
  "visual_style": "Describe the visual style: live-action or animation style (2D/3D/CGI/hand-drawn/anime/mixed), color palette, lighting mood, any notable visual effects or stylistic choices."
}

CRITICAL RULES
1) ENGLISH ONLY.
2) Output must be parseable JSON (no trailing commas, no markdown, no code fences).
3) "caption" --- VISUAL ONLY, a SINGLE coherent multi-sentence paragraph (NOT a list).
   - If multiple shots: describe each shot sequentially, clearly marking transitions.
   - If single shot: describe chronological progression of action.
   - For each shot/moment: framing (wide/medium/close-up), camera angle, camera motion, what is visible, what changes.
   - Mention on-screen text/titles/subtitles if clearly readable.
   - Do NOT describe any audio content in this field.
4) "audio_caption" --- AUDIO ONLY, a SINGLE coherent multi-sentence paragraph covering ALL sound.
   - Speech: describe every instance of speech chronologically. For each speaker: visual appearance if on-screen (hair, clothing, gender presentation), or "off-screen voice" / "narrator". Describe tone (calm/excited/angry/sad/whispering/shouting), emotion, pace (fast/slow/measured), volume, and speaking style (dialogue/monologue/narration/voiceover). Do NOT transcribe or quote any spoken words.
   - Sound effects: describe sounds tied to on-screen events (footsteps, impacts, glass breaking, door closing, object falling, water splashing, engine roar, etc.) and when they occur relative to the clip.
   - Ambient / environmental sounds: room tone, crowd murmur, wind, rain, traffic, birdsong, silence, etc.
   - Background music: genre, mood (tense/uplifting/melancholic/epic), instrumentation (strings/piano/synth/drums/choir), tempo (BPM range or qualitative), intensity, and how it evolves across the clip (builds, fades, shifts mood/key, stops abruptly, etc.).
   - Describe how these layers interact: does music swell when speech pauses? Does a sound effect punctuate a transition? Does ambient sound shift between shots?
   - If audio is completely silent, write "silence throughout".
5) "shot_count": integer, how many distinct shots/cuts you observe (1 for single-shot video).
6) "speech_or_non_speech": "speech" if ANY human speech is audible, otherwise "non_speech".
7) "visual_style": describe the artistic/rendering style, not plot or action.

DETAIL CHECKLIST for "caption" (VISUAL, include as many as clearly verifiable per shot):
1) Setting & Background:
   - indoor/outdoor, location type, major objects, signage/text if readable,
     lighting, time-of-day cues, weather if visible.
2) Characters / Subjects:
   - character/person descriptions (hair, clothing, distinguishing features),
     count, facial expressions, gaze direction, body posture.
3) Actions & Interactions:
   - what subjects do, gestures, movement direction,
     interactions between subjects or with objects,
     key changes within each shot.
4) Frame / Camera:
   - shot type (establishing/wide/medium/close-up/extreme close-up),
     camera angle (eye-level/high/low/bird's-eye/dutch),
     camera motion (static/pan/tilt/zoom/tracking/dolly),
     depth-of-field if notable.
5) Transitions (for multi-shot only):
   - cut type if notable (hard cut/dissolve/fade/wipe/match cut),
     pacing of cuts (fast/slow/rhythmic).

DETAIL CHECKLIST for "audio_caption" (include as many as clearly audible):
1) Speech:
   - who speaks (visual description or off-screen), tone, emotion, pace, volume,
     language if identifiable, dialogue/monologue/narration/voiceover,
     approximate number of speakers, turn-taking patterns.
2) Sound Effects:
   - source object/action, loudness, timing relative to visual events,
     spatial quality (near/far, left/right if notable).
3) Ambient / Environment:
   - room tone, outdoor atmosphere, crowd presence, weather sounds,
     changes between shots or scenes.
4) Music:
   - genre, mood, instrumentation, tempo, dynamics (loud/soft/crescendo/decrescendo),
     how it starts/stops/transitions, relationship to on-screen action.

ANTI-REDUNDANCY
- Do NOT repeat the same details across shots unless something clearly changes.
- Each shot description should focus on what is NEW or DIFFERENT from the previous shot.
- Avoid generic filler like "the scene continues" --- be specific about what happens.

FINAL EMPHASIS
Be as detailed as possible while staying 100% factual. "caption" and "audio_caption" must be independent --- each should be fully understandable on its own without reading the other.
\end{lstlisting}

\subsection{Checklist Generation Prompts}
\label{app:bench-prompts}

The following four V2 generators create modality-specific, source-grounded checklist questions for video-only, audio-only, joint audio-video, and speech editing tasks.

\subsubsection{Video-Only Checklist Generator}
\label{app:bench-prompt-video}
\begin{lstlisting}[style=promptstyle]
CHECKLIST_GENERATOR_SYSTEM_PROMPT = """# Role

You generate a V2 checklist for one VIDEO-ONLY edit.

The requested edit changes visible content only.

You MUST NOT evaluate media.
You MUST NOT answer the checklist.
You MUST output valid JSON only.

# Output Spec

Return:
{
  "schema_version": "checklist_v2",
  "edit_category": "video_only",
  "questions": [
    {
      "question_id": "Q01",
      "dimension": "Edit Response",
      "subdimension": "edit_response",
      "modality_tag": "video",
      "question": "..."
    }
  ]
}

Allowed dimensions:
- "Edit Response"
- "Instruction Following"
- "Fidelity"

Allowed subdimensions:
- "edit_response"
- "instruction_correctness"
- "non_edit_preservation"
- "hallucination_control"

Allowed modality_tag values: "video", "audio", "general".

# Answer Polarity

Phrase EVERY question so that "Yes" is the desired/positive answer (edit done correctly, element preserved, or NO unrequested addition). NEVER phrase a question so that "Yes" indicates a defect, error, or hallucination. For hallucination-control, always use the "no extra / free of" form, e.g. "Is the target audio free of any unrequested added music/voiceover/sound effects?" --- NOT "Does the target audio contain extra audio?".

# Required Structure

Generate a concise checklist, typically 8-16 questions. The question count is
not fixed.

Q01 must be the only Edit Response question:
"Does the target video contain any visible change compared with the source video?"

Do NOT generate an audio response question for video-only edits.

Q01 is descriptive response-rate only. It does not ask whether the visible
change is correct, natural, or faithful.

# Instruction Following

Generate AT MOST 6 instruction-following questions; keep only the most important ones.

Generate graded visual correctness questions:
- partial visual correctness: use a lower-threshold question such as whether a
  visible change in the requested direction appears at all;
- full visual correctness / degree: use a stronger question such as whether the
  requested visual edit clearly reaches the intended object/action/degree;
- requested object/action/category correctness;
- requested attributes such as color, quantity, position, size, speed, or
  intensity;
- whether the intended subject/region is edited rather than the wrong target;
- coverage duration for persistent visual edits such as background/environment
  changes, identity/appearance changes, object replacement/removal while the
  object is visible, or explicit full-span requirements;
- degree/action-completion questions for action-type edits where the requested
  change is a temporal action or event rather than a persistent state;
- original target absence for replacement/removal when applicable.

Partial and full correctness questions must be meaningfully different. Do not
ask duplicate pairs that only rephrase the same threshold.

Do not use degree questions as a substitute for coverage on persistent visual
background/environment/identity/appearance/object edits. For action-type edits,
use degree/action-completion questions instead of default coverage questions.

For replacement edits, separate "new target appears" from "old target
disappears". Do not combine the new target and old-target absence in the same
question.

Do not include audio instruction-following questions unless the prompt itself
explicitly contains an audio requirement. For pure video-only edits, scored
Instruction Following should be video-tagged.

# Fidelity

The Fidelity dimension must contain AT MOST 7 questions in total (non_edit_preservation + hallucination_control combined); prioritize the most important retained elements.

Generate source-vs-target preservation questions. Each preservation question
must name the concrete properties to check, so the judging criteria live in the
question wording itself:
- non-edited visual elements --- ask whether the element is preserved with no
  significant change in shape, texture, appearance/color, or size;
- camera movement, framing behavior, shot cuts/transitions --- if the source has
  ANY camera movement OR any shot cut/transition, you MUST include a question
  asking whether the camera movement and shot cuts/transitions match the source;
- moving-object trajectories if relevant;
- retained audio only if the source audio is meaningful and should remain
  unchanged --- ONLY when MULTIPLE distinct non-edited audio sources are present,
  generate a SEPARATE preservation question for EACH (foreground non-edited sounds;
  background music; narration/voiceover/retained speech; ambient sound); if there
  is only ONE retained source, a single question suffices. Each asks whether it is
  preserved with no significant change in character, level, pitch, or content. When
  the source has people speaking and that speech is retained (not the edit target),
  split the retained-speech preservation into TWO questions: one on whether the
  spoken content/words are preserved, and one on whether the speaker's voice
  tone/timbre/pitch is preserved.

Phrase each question so that a clearly noticeable change in its named properties
reads as "No", while minor or imperceptible differences read as "Yes".

Camera drift isolation: mild global framing/crop/scale/translation drift should
be assigned to the camera/framing question, not every background/object/person
question.

Generate hallucination-control questions: no extra visual objects/subtitles/text/
logos/watermarks or new unrequested events; and if the source audio should remain
unchanged, a separate audio hallucination question (no extra voiceover/BGM/sound
effects/unrelated audio).

# Atomicity

Every scored question after Q01 must reference concrete nouns, actions, visual
elements, or events from the Source Video Caption or Edit Prompt.

Each question asks exactly one observable thing.
"""

CHECKLIST_GENERATOR_USER_PROMPT = """Source Video Caption:
{caption}

Edit Prompt:
{edit_prompt}

Edit Category:
video_only

Generate a V2 video-only checklist as valid JSON only."""
\end{lstlisting}

\subsubsection{Audio-Only Checklist Generator}
\label{app:bench-prompt-audio}
\begin{lstlisting}[style=promptstyle]
CHECKLIST_GENERATOR_SYSTEM_PROMPT = """# Role

You generate a V2 checklist for one AUDIO-ONLY edit.

The requested edit changes audible content only.

You MUST NOT evaluate media.
You MUST NOT answer the checklist.
You MUST output valid JSON only.

# Output Spec

Return:
{
  "schema_version": "checklist_v2",
  "edit_category": "audio_only",
  "questions": [
    {
      "question_id": "Q01",
      "dimension": "Edit Response",
      "subdimension": "edit_response",
      "modality_tag": "audio",
      "question": "..."
    }
  ]
}

Allowed dimensions:
- "Edit Response"
- "Instruction Following"
- "Fidelity"

Allowed subdimensions:
- "edit_response"
- "instruction_correctness"
- "non_edit_preservation"
- "hallucination_control"

Allowed modality_tag values: "video", "audio", "general".

# Answer Polarity

Phrase EVERY question so that "Yes" is the desired/positive answer (edit done correctly, element preserved, or NO unrequested addition). NEVER phrase a question so that "Yes" indicates a defect, error, or hallucination. For hallucination-control, always use the "no extra / free of" form, e.g. "Is the target audio free of any unrequested added music/voiceover/sound effects?" --- NOT "Does the target audio contain extra audio?".

# Required Structure

Generate a concise checklist, typically 8-16 questions. The question count is
not fixed.

Q01 must be the only Edit Response question:
"Does the target audio contain any audible change compared with the source audio?"

Do NOT generate a video response question for audio-only edits.

Q01 is descriptive response-rate only. It does not ask whether the audible
change is correct, natural, or faithful.

# Instruction Following

Generate AT MOST 6 instruction-following questions; keep only the most important ones.

Generate graded audio correctness questions:
- partial audio correctness: use a lower-threshold question such as whether an
  audible change in the requested direction appears at all;
- full audio correctness / degree: use a stronger question such as whether the
  requested audio edit clearly reaches the intended sound/voice/degree;
- requested sound/speech/music/category correctness;
- requested attributes such as loudness, intensity, rhythm, pitch, timbre,
  language, speaker attribute, or sound character;
- coverage duration for persistent audio edits such as ambient sound changes,
  voice/speech changes, retained/replaced continuous sounds, or explicit
  full-span requirements;
- degree/action-completion questions for action-type sounds where the requested
  change is a temporal event rather than a persistent state;
- original sound absence for replacement/removal when applicable.

Partial and full correctness questions must be meaningfully different. Do not
ask duplicate pairs that only rephrase the same threshold.

Do not use degree questions as a substitute for coverage on persistent ambient,
voice/speech, retained, or replaced continuous audio edits. For action-type
sounds, use degree/action-completion questions instead of default coverage
questions.

For replacement edits, separate "new target appears" from "old target
disappears". Do not combine the new target and old-target absence in the same
question.

Do not include video instruction-following questions unless the prompt itself
explicitly contains a visual requirement. For pure audio-only edits, scored
Instruction Following should be audio-tagged.

# Fidelity

The Fidelity dimension must contain AT MOST 7 questions in total (non_edit_preservation + hallucination_control combined); prioritize the most important retained elements.

Generate source-vs-target preservation questions. Each preservation question
must name the concrete properties to check, so the judging criteria live in the
question wording itself:
- retained non-edited audio --- ONLY when MULTIPLE distinct non-edited audio
  sources are present, generate a SEPARATE preservation question for EACH and do
  NOT lump them together; if there is only ONE retained source, a single question
  suffices. Sources to distinguish, for whichever are present: (a) foreground
  non-edited sounds (retained action sounds / sound effects); (b) background music
  (BGM); (c) narration/voiceover/retained speech; (d) ambient/background sound.
  Each asks whether that source is preserved with no significant change in
  character, level, pitch, or content. When the source has people speaking and
  that speech is retained (not the edit target), split source (c) into TWO
  questions: one on whether the spoken content/words are preserved, and one on
  whether the speaker's voice tone/timbre/pitch is preserved;
- visual/camera preservation only if video is present and should remain
  unchanged --- ask whether non-edited visual elements are preserved with no
  significant change in shape, texture, appearance/color, or size; and if the
  source has ANY camera movement OR shot cut/transition, you MUST include a
  question asking whether the camera movement and shot cuts/transitions match the
  source.

Phrase each question so that a clearly noticeable change in its named properties
reads as "No", while minor or imperceptible differences read as "Yes".

Generate hallucination-control questions:
- no extra voiceover/BGM/sound effects/speaker turns/unrelated audio (a separate
  audio hallucination question);
- no new unrequested audible events;
- include visual hallucination only if video is present.

# Atomicity

Every scored question after Q01 must reference concrete sounds, speech, music,
speakers, or events from the Source Video Caption or Edit Prompt.

Each question asks exactly one observable thing.
"""

CHECKLIST_GENERATOR_USER_PROMPT = """Source Video Caption:
{caption}

Edit Prompt:
{edit_prompt}

Edit Category:
audio_only

Generate a V2 audio-only checklist as valid JSON only."""
\end{lstlisting}

\subsubsection{Joint Audio-Video Checklist Generator}
\label{app:bench-prompt-joint}
\begin{lstlisting}[style=promptstyle]
CHECKLIST_GENERATOR_SYSTEM_PROMPT = """# Role

You generate a V2 checklist for one JOINT audio-video edit.

The edit changes both video and audio, or the requested sound must correspond
to a visible event.

You MUST NOT evaluate media.
You MUST NOT answer the checklist.
You MUST output valid JSON only.

# Output Spec

Return:
{
  "schema_version": "checklist_v2",
  "edit_category": "joint",
  "questions": [
    {
      "question_id": "Q01",
      "dimension": "Edit Response",
      "subdimension": "edit_response",
      "modality_tag": "video",
      "question": "..."
    }
  ]
}

Allowed dimensions:
- "Edit Response"
- "Instruction Following"
- "Fidelity"

Allowed subdimensions:
- "edit_response"
- "instruction_correctness"
- "non_edit_preservation"
- "hallucination_control"

Allowed modality_tag values: "video", "audio", "general".

# Answer Polarity

Phrase EVERY question so that "Yes" is the desired/positive answer (edit done correctly, element preserved, or NO unrequested addition). NEVER phrase a question so that "Yes" indicates a defect, error, or hallucination. For hallucination-control, always use the "no extra / free of" form, e.g. "Is the target audio free of any unrequested added music/voiceover/sound effects?" --- NOT "Does the target audio contain extra audio?".

# Required Structure

Generate a concise checklist, typically 10-18 questions. The question count is
not fixed.

Suggested layout:
- Q01-Q02: Edit Response / edit_response
- next questions: Instruction Following / instruction_correctness
- next questions: Fidelity / non_edit_preservation
- final questions: Fidelity / hallucination_control

Q01 must be:
"Does the target video contain any visible change compared with the source video?"

Q02 must be:
"Does the target audio contain any audible change compared with the source audio?"

Q01-Q02 are descriptive response-rate questions only. They do not ask whether
the change is correct, natural, or faithful.

# Instruction Following

Generate AT MOST 6 instruction-following questions; if more checks are possible,
keep only the most important ones.

Generate graded correctness questions, not simple response questions.

Cover:
- partial visual correctness: use a lower-threshold question such as whether a
  visible change in the requested direction appears at all;
- full visual correctness / degree: use a stronger question such as whether the
  requested visual edit clearly reaches the intended object/action/degree;
- partial audio correctness: use a lower-threshold question such as whether an
  audible change in the requested direction appears at all;
- full audio correctness / degree: use a stronger question such as whether the
  requested audio edit clearly reaches the intended sound/voice/degree;
- requested visual/audio attributes, degree, intensity, quantity, timing, or
  sound character;
- visual coverage duration for persistent visual edits such as background or
  environment changes, identity/appearance changes, object replacement/removal
  while the object is visible, or explicit full-span requirements;
- audio coverage duration for persistent audio edits such as ambient sound
  changes, voice/speech changes, retained/replaced continuous sounds, or
  explicit full-span requirements;
- degree/action-completion questions for action-type edits where the requested
  change is a temporal action or event rather than a persistent state;
- original target absence for replacement/removal when applicable.

Do not ask about naturalness, artifacts, sync, source preservation, or
hallucinations in this dimension.

Do not ask a generic overall joint-completion question such as whether the video
and audio edit are "completed together". If coverage matters, split it into one
video coverage question and one audio coverage question.

Do not use degree questions as a substitute for coverage on persistent
background/environment/ambient/identity/object/voice edits. For action-type
edits, use degree/action-completion questions instead of default coverage
questions.

Partial and full correctness questions must be meaningfully different. Do not
ask duplicate pairs that only rephrase the same threshold.

For replacement edits, separate "new target appears" from "old target
disappears". Do not combine the new target and old-target absence in the same
question. For example:
- Good: "Does the target audio contain a dull earth thud?"
- Good: "Is the original soft snow-lifting sound absent?"
- Bad: "Does the target audio contain gritty scraping and replace the original
  soft snow-lifting sound?"

# Fidelity

The Fidelity dimension must contain AT MOST 7 questions in total
(non_edit_preservation + hallucination_control combined). Prioritize the most
important retained elements and drop the rest to stay within the limit.

Generate source-vs-target preservation questions for non-edited content. Each
preservation question must name the concrete properties to check, so the
judging criteria live in the question wording itself:
- main non-edited visual elements --- ask whether the element is preserved with no
  significant change in shape, texture, appearance/color, or size (e.g. "Is the
  wooden table preserved with the same shape, texture, color, and size as the
  source?");
- retained audio --- ONLY when the source contains MULTIPLE distinct non-edited
  audio sources, generate a SEPARATE preservation question for EACH of them and do
  NOT lump them together; if there is only ONE retained source, a single question
  suffices. Sources to distinguish, for whichever are present:
  (a) foreground non-edited sounds (retained action sounds / sound effects);
  (b) background music (BGM);
  (c) narration / voiceover / retained speech;
  (d) ambient / background sound.
  Each asks whether that source is preserved with no significant change in its
  character, level, pitch, or content. When the source has people speaking and
  that speech is retained (not the edit target), split source (c) into TWO
  questions: one asking whether the spoken content/words are preserved, and one
  asking whether the speaker's voice tone/timbre/pitch is preserved;
- camera movement, framing behavior, shot cuts/transitions --- if the source has
  ANY camera movement OR any shot cut/transition, you MUST include a question
  asking whether the camera movement and shot cuts/transitions match the source;
- moving-object trajectories if relevant.

Phrase each question so that a clearly noticeable change in its named properties
reads as "No", while minor or imperceptible differences read as "Yes".

Camera drift isolation: mild global framing/crop/scale/translation drift should
be assigned to the camera/framing question, not every background/object/person
question.

Generate hallucination-control questions:
- no extra visual objects/subtitles/text/logos/watermarks;
- no extra voiceover/BGM/sound effects/unrelated audio (a separate audio
  hallucination question);
- no new unrequested events.

# Atomicity

Every scored question after Q01-Q02 must reference concrete nouns, actions,
visual elements, sounds, speakers, or events from the Source Video Caption or
Edit Prompt.

Each question asks exactly one observable thing.
"""

CHECKLIST_GENERATOR_USER_PROMPT = """Source Video Caption:
{caption}

Edit Prompt:
{edit_prompt}

Edit Category:
joint

Generate a V2 joint-edit checklist as valid JSON only."""
\end{lstlisting}

\subsubsection{Speech Checklist Generator}
\label{app:bench-prompt-speech}
\begin{lstlisting}[style=promptstyle]
CHECKLIST_GENERATOR_SYSTEM_PROMPT = """# Role

You generate a V2 checklist for one SPEECH edit.

The requested edit changes speech, voice, dialogue, speaker attributes,
language, or spoken content.

You MUST NOT evaluate media.
You MUST NOT answer the checklist.
You MUST output valid JSON only.

# Output Spec

Return:
{
  "schema_version": "checklist_v2",
  "edit_category": "speech",
  "questions": [
    {
      "question_id": "Q01",
      "dimension": "Edit Response",
      "subdimension": "edit_response",
      "modality_tag": "audio",
      "question": "..."
    }
  ]
}

Allowed dimensions:
- "Edit Response"
- "Instruction Following"
- "Fidelity"

Allowed subdimensions:
- "edit_response"
- "instruction_correctness"
- "non_edit_preservation"
- "hallucination_control"

Allowed modality_tag values: "video", "audio", "general".

# Answer Polarity

Phrase EVERY question so that "Yes" is the desired/positive answer (edit done correctly, element preserved, or NO unrequested addition). NEVER phrase a question so that "Yes" indicates a defect, error, or hallucination. For hallucination-control, always use the "no extra / free of" form, e.g. "Is the target audio free of any unrequested added music/voiceover/sound effects?" --- NOT "Does the target audio contain extra audio?".

# Required Structure

Generate a concise checklist, typically 8-16 questions. The question count is
not fixed.

Q01 must be audio response:
"Does the target audio contain any audible change compared with the source audio?"

Add a video response question only if the speech edit is expected to affect
visible mouth/face motion, subtitles, speaker identity, or other visual content.

Edit Response questions are descriptive response-rate only. They do not ask
whether the change is correct, natural, or faithful.

# Instruction Following

Generate AT MOST 6 instruction-following questions; keep only the most important ones.

Generate graded speech correctness questions:
- partial speech/voice correctness: requested speech/voice/dialogue appears in
  the right semantic direction;
- full speech/voice correctness / degree: the requested words, language,
  speaker attribute, or voice change clearly reaches the intended result;
- spoken words/content/language are correct;
- requested speaker attribute, voice identity, gender presentation, tone, or
  style is correct;
- coverage duration for voice/speech changes over the relevant speech span, or
  explicit full-span requirements;
- degree/action-completion questions only when the speech edit changes a
  temporal speaking action/event rather than a persistent voice/content state;
- original speech/voice absence for replacement/removal when applicable.

Partial and full correctness questions must be meaningfully different. Do not
ask duplicate pairs that only rephrase the same threshold.

Do not use degree questions as a substitute for coverage on voice/content
replacement edits. Use degree/action-completion questions only for action-type
speech edits.

For replacement edits, separate "new target appears" from "old target
disappears". Do not combine the new target and old-target absence in the same
question.

Do not infer voice gender, speaker identity, or spoken content from video
evidence. Use audible evidence for audio questions.

# Fidelity

The Fidelity dimension must contain AT MOST 7 questions in total (non_edit_preservation + hallucination_control combined); prioritize the most important retained elements.

Generate source-vs-target preservation questions. Each preservation question
must name the concrete properties to check, so the judging criteria live in the
question wording itself:
- retained non-edited speech timing/content when applicable --- ask whether it is
  preserved with no significant change in content, timing, or pacing;
- retained speaker tone/timbre where it should remain --- ask whether it is
  preserved with no significant change in character, level, or pitch;
- non-target audio --- ONLY when MULTIPLE distinct non-edited audio sources are
  present, generate a SEPARATE preservation question for EACH (background music;
  ambient/background sound; other foreground non-edited sounds); if there is only
  ONE retained source, a single question suffices. Each asks whether it is
  preserved with no significant change in character, level, or content;
- visual identity, face/mouth region when not intentionally edited --- ask whether
  they are preserved with no significant change in shape, texture, appearance/
  color, or size; and if the source has ANY camera movement OR shot cut/
  transition, you MUST include a question asking whether the camera movement and
  shot cuts/transitions match the source.

Phrase each question so that a clearly noticeable change in its named properties
reads as "No", while minor or imperceptible differences read as "Yes".

Generate hallucination-control questions:
- no extra spoken words or extra speaker turns beyond the requested edit;
- no extra voiceover/BGM/sound effects/unrelated audio (a separate audio
  hallucination question);
- no extra subtitles/on-screen text or visual additions (if video is present).

# Atomicity

Every scored question after response questions must reference concrete speech,
voice, speaker, audio, visual, or event details from the Source Video Caption or
Edit Prompt.

Each question asks exactly one observable thing.
"""

CHECKLIST_GENERATOR_USER_PROMPT = """Source Video Caption:
{caption}

Edit Prompt:
{edit_prompt}

Edit Category:
speech

Generate a V2 speech checklist as valid JSON only."""
\end{lstlisting}
\subsection{Checklist Evaluation Prompts}
\label{app:bench-checklist-prompts}

Checklist evaluation uses three evaluator system prompts plus split user prompts for audio-only, video-only, and general audio-video checklist items. These prompts instruct the evaluator to answer each checklist item from direct audiovisual evidence while preserving the checklist schema.

\subsubsection{Video-Only Checklist Evaluator}
\label{app:prompt-eval-video}
\begin{lstlisting}[style=promptstyle]
CHECKLIST_EVALUATOR_SYSTEM_PROMPT = """# Role
Evaluate a target video-only edit against a fixed V2 checklist (source audio should stay unchanged). Answer every item Yes or No from direct evidence. Do not modify, add, or remove questions. Output valid JSON only.

# Answering Principle
- Use only directly observable/audible evidence.
- "Yes" only when the question's condition is clearly met; otherwise "No" (absent, ambiguous, partial, weak, or only approximate).
- Do not infer success from the edit prompt. No credit for near-misses.

# Dimensions

## Edit Response --- did the video change vs source?
Descriptive, not correctness. Strict, low threshold: ANY visible change = Yes; "No" only if the target video is essentially identical to source. Do not judge whether the change is correct or natural.

## Instruction Following --- was the requested edit done correctly?
"Yes" only if the requested semantics / attribute / degree / coverage / removal is clearly satisfied.
"No" if: wrong object/action/category/color/quantity/position/size/speed/intensity; wrong subject; present only briefly when full coverage required; partial when completion required; or original target still present when removal required.
Do not judge: whether anything changed (Edit Response), naturalness (Realism), source preservation (Fidelity), or extra content (hallucination_control).

## Fidelity / non_edit_preservation --- did non-edited parts stay the same?
Source-vs-target, by overall perception NOT pixel-by-pixel; minor/imperceptible differences = Yes.
Each question names the element and the properties to check --- judge exactly that: a clearly noticeable change in the named property = No; trivial difference = Yes. Do not check the edited target itself here.
Edit-consequence (applies to BOTH video and audio): ignore changes to a non-edited element that are a necessary/direct consequence of the requested edit --- answer Yes. For video, e.g. shadows, reflections, lighting, occlusion caused by the edited target; for audio, e.g. masking, mixing, or level changes of retained sounds caused by the edited sound. Only count changes unrelated to the edit.
Camera drift: a single uniform global framing/scale/position offset -> mark only the camera/framing question No, not every element; it does not excuse a real local change.
Aspect-ratio / frame completion: ignore any change caused by a different aspect ratio or by outpainting/padding that fills newly exposed areas. Do not answer "No" because of the aspect-ratio change itself or anything in the filled-in regions; judge only the originally-visible content.
If a retained-audio question is present, "No" when the audio element is significantly altered in character/level/pitch/content.

## Fidelity / hallucination_control --- any unrequested additions?
"No" for: extra visual objects/subtitles/text/logos/watermarks; new unrequested visual events; and --- for audio (if a retained-audio hallucination question is present) --- penalize ONLY clearly NEW added audio: added voiceover/narration, added background music, or a significant new independent event sound. Do NOT penalize leftover ORIGINAL audio that was not fully removed (e.g., partially retained voice or residual background noise from the source), nor minor/faint noise or low-level artifacts. Content that appears only because of an aspect-ratio change/frame completion, or any sound that is a direct consequence of the requested edit, is NOT an unrequested addition --- ignore it.

# Gate (applied later in stats; still answer every item)
Video response No -> drop video fidelity/realism.

# Output
JSON only; keep each item's question_id / dimension / subdimension / modality_tag / question unchanged:
{
  "schema_version": "checklist_v2",
  "edit_category": "video_only",
  "visual_discrepancy_analysis": "short comparison summary",
  "evaluations": [
    {"question_id": "Q01", "dimension": "Edit Response", "subdimension": "edit_response", "modality_tag": "video", "question": "...", "observation": "short evidence", "answer": "Yes", "justification": "short reason"}
  ],
  "summary": {"total_questions": 0, "yes_count": 0, "no_count": 0, "score": 0.0}
}
"""
\end{lstlisting}

\subsubsection{Audio-Only Checklist Evaluator}
\label{app:prompt-eval-audio}
\begin{lstlisting}[style=promptstyle]
CHECKLIST_EVALUATOR_SYSTEM_PROMPT = """# Role
Evaluate a target audio-only edit against a fixed V2 checklist (source video should stay unchanged). Answer every item Yes or No from direct evidence. Do not modify, add, or remove questions. Output valid JSON only.

# Answering Principle
- Use only directly observable/audible evidence.
- "Yes" only when the question's condition is clearly met; otherwise "No" (absent, ambiguous, partial, weak, or only approximate).
- Do not infer success from the edit prompt. No credit for near-misses.

# Dimensions

## Edit Response --- did the audio change vs source?
Descriptive, not correctness. Strict, low threshold: ANY audible change = Yes; "No" only if the target audio is essentially identical to source. Do not judge whether the change is correct or natural.

## Instruction Following --- was the requested edit done correctly?
"Yes" only if the requested semantics / attribute / degree / coverage / removal is clearly satisfied.
"No" if: wrong sound/voice/language/category; wrong loudness/intensity/rhythm/pitch/timbre/speaker/content; wrong target; present only briefly when full coverage required; partial when completion required; or original sound still present when removal required.
Do not judge: whether anything changed (Edit Response), naturalness/sync (Realism), source preservation (Fidelity), or extra content (hallucination_control).

## Fidelity / non_edit_preservation --- did non-edited parts stay the same?
Source-vs-target, by overall perception NOT sample-by-sample; minor/imperceptible differences = Yes.
Each question names the element and the properties to check --- judge exactly that: a clearly noticeable change in the named property = No; trivial difference = Yes. Do not check the edited target itself here.
Edit-consequence (applies to BOTH audio and video): ignore changes to a non-edited element that are a necessary/direct consequence of the requested edit --- answer Yes. For audio, e.g. masking, mixing, or level changes of retained sounds caused by the edited sound; for video, e.g. shadows/lighting/occlusion caused by the edited target. Only count changes unrelated to the edit.
For retained audio (timbre/tone/ambience/music/SFX/retained speech), "No" when significantly altered in character/level/pitch/content.
If a visual-preservation question is present, "No" only when a non-edited visual element changes significantly in shape, texture, appearance/color, or size. Ignore any change caused by a different aspect ratio or by frame completion/outpainting --- judge only the originally-visible content.

## Fidelity / hallucination_control --- any unrequested additions?
Penalize ONLY clearly NEW added audio: "No" for added voiceover/narration, added background music, a significant new independent event sound, or extra speaker turns; and extra visual objects/subtitles/text (if a visual hallucination question is present). Do NOT penalize leftover ORIGINAL audio that was not fully removed (e.g., partially retained voice or residual background noise from the source), nor minor/faint noise or low-level artifacts. Any sound that is a direct consequence of the requested edit is NOT an unrequested addition --- ignore it.

# Gate (applied later in stats; still answer every item)
Audio response No -> drop audio fidelity/realism/quality.

# Output
JSON only; keep each item's question_id / dimension / subdimension / modality_tag / question unchanged:
{
  "schema_version": "checklist_v2",
  "edit_category": "audio_only",
  "visual_discrepancy_analysis": "short comparison summary",
  "evaluations": [
    {"question_id": "Q01", "dimension": "Edit Response", "subdimension": "edit_response", "modality_tag": "audio", "question": "...", "observation": "short evidence", "answer": "Yes", "justification": "short reason"}
  ],
  "summary": {"total_questions": 0, "yes_count": 0, "no_count": 0, "score": 0.0}
}
"""
\end{lstlisting}

\subsubsection{Restructured Joint/Speech Checklist Evaluator}
\label{app:prompt-eval-restructured}
\begin{lstlisting}[style=promptstyle]
CHECKLIST_EVALUATOR_SYSTEM_PROMPT = """# Role
Evaluate a target output (joint or speech edit) against a fixed V2 checklist. Answer every item Yes or No from direct evidence. Do not modify, add, or remove questions. Output valid JSON only.

# Answering Principle
- Use only directly observable/audible evidence.
- "Yes" only when the question's condition is clearly met; otherwise "No" (absent, ambiguous, partial, weak, or only approximate).
- Do not infer success from the edit prompt. No credit for near-misses.

# Dimensions

## Edit Response --- did this modality change vs source?
Descriptive, not correctness. Strict, low threshold: ANY detectable change = Yes; "No" only if essentially identical to source. Do not judge whether the change is correct or natural.
- Video response: compare source vs target video.
- Audio response: compare source vs target audio.
- Speech: audio response expected; include video response only if the edit affects visible mouth/face/subtitles.

## Instruction Following --- was the requested edit done correctly?
"Yes" only if the requested semantics / attribute / degree / coverage / removal is clearly satisfied.
"No" if: wrong object/action/sound/voice/language/color/quantity/speed/intensity/content; wrong subject; present only briefly when full coverage required; partial when completion required; or original target still present when removal required.
Do not judge: whether anything changed (Edit Response), naturalness/sync (Realism), source preservation (Fidelity), or extra content (hallucination_control).

## Fidelity / non_edit_preservation --- did non-edited parts stay the same?
Source-vs-target, by overall perception NOT pixel-by-pixel; minor/imperceptible differences = Yes.
Each question names the element and the properties to check --- judge exactly that: a clearly noticeable change in the named property = No; trivial difference = Yes. Do not check the edited target itself here.
Edit-consequence (applies to BOTH video and audio): ignore changes to a non-edited element that are a necessary/direct consequence of the requested edit --- answer Yes. For video, e.g. shadows, reflections, lighting, occlusion caused by the edited target; for audio, e.g. masking, mixing, or level changes of retained sounds caused by the edited sound. Only count changes unrelated to the edit.
Camera drift: a single uniform global framing/scale/position offset -> mark only the camera/framing question No, not every element; it does not excuse a real local change.
Aspect-ratio / frame completion: ignore any change caused by a different aspect ratio or by outpainting/padding that fills newly exposed areas. Do not answer "No" because of the aspect-ratio change itself or anything in the filled-in regions; judge only the originally-visible content.

## Fidelity / hallucination_control --- any unrequested additions?
"No" for: extra visual objects/subtitles/text/logos/watermarks; extra spoken words or speaker turns; new unrequested events; and --- for audio --- penalize ONLY clearly NEW added audio: added voiceover/narration, added background music, or a significant new independent event sound. Do NOT penalize leftover ORIGINAL audio that was not fully removed (e.g., partially retained voice or residual background noise from the source), nor minor/faint noise or low-level artifacts. Content that appears only because of an aspect-ratio change/frame completion, or any sound that is a direct consequence of the requested edit, is NOT an unrequested addition --- ignore it.

# Gate (applied later in stats; still answer every item)
Video response No -> drop video fidelity/realism. Audio response No -> drop audio fidelity/realism/quality. Both No -> drop AV-sync.

# Output
JSON only; keep each item's question_id / dimension / subdimension / modality_tag / question unchanged:
{
  "schema_version": "checklist_v2",
  "edit_category": "joint",
  "visual_discrepancy_analysis": "short comparison summary",
  "evaluations": [
    {"question_id": "Q01", "dimension": "Edit Response", "subdimension": "edit_response", "modality_tag": "video", "question": "...", "observation": "short evidence", "answer": "Yes", "justification": "short reason"}
  ],
  "summary": {"total_questions": 0, "yes_count": 0, "no_count": 0, "score": 0.0}
}
"""
\end{lstlisting}

\subsubsection{Split Evaluation User Prompts}
\label{app:prompt-eval-split-user}
\begin{lstlisting}[style=promptstyle]
# Split Evaluation User Prompts

These templates are used as user prompts during checklist evaluation. The evaluator splits checklist items by `modality_tag` and sends separate calls for audio-only, video-only, and general A/V questions.

## Audio-Only

```text
# Evaluation Input (Audio-Only)

You are provided with SOURCE audio and TARGET audio only. No video frames are provided.

Source Video Caption:
{source_caption}

Edit Prompt:
{edit_prompt}

Edit Type:
{edit_type}

Checklist subset (audio questions only):
{checklist_text}

Instructions:
1. Answer every checklist item above.
2. Keep question_id, question, dimension, subdimension, and modality_tag unchanged.
3. Base judgments only on audible evidence from source/target audio.
4. Judge each dimension by the rules in the system prompt; for Fidelity, always compare source and target audio.

Respond in JSON:
{
  "visual_discrepancy_analysis": "short audio-focused discrepancy analysis",
  "evaluations": [
    {
      "question_id": "Q01",
      "dimension": "Instruction Following",
      "subdimension": "instruction_correctness",
      "modality_tag": "audio",
      "question": "...",
      "observation": "...",
      "answer": "Yes/No",
      "justification": "..."
    }
  ],
  "summary": {"total_questions": {len(checklist_items)}, "yes_count": 0, "no_count": 0, "score": 0.0}
}
```

## Video-Only

```text
# Evaluation Input (Video-Only)

You are provided with SOURCE video and TARGET video only. No standalone audio files are provided.

Source Video Caption:
{source_caption}

Edit Prompt:
{edit_prompt}

Edit Type:
{edit_type}

Checklist subset (video questions only):
{checklist_text}

Instructions:
1. Answer every checklist item above.
2. Keep question_id, question, dimension, subdimension, and modality_tag unchanged.
3. Base judgments only on visual evidence from source/target video.
4. For Instruction Following state questions, target-only evaluation is allowed.
5. Judge each dimension by the rules in the system prompt; for Fidelity, always compare source and target video.

Respond in JSON:
{
  "visual_discrepancy_analysis": "short visual-focused discrepancy analysis",
  "evaluations": [
    {
      "question_id": "Q01",
      "dimension": "Instruction Following",
      "subdimension": "instruction_correctness",
      "modality_tag": "video",
      "question": "...",
      "observation": "...",
      "answer": "Yes/No",
      "justification": "..."
    }
  ],
  "summary": {"total_questions": {len(checklist_items)}, "yes_count": 0, "no_count": 0, "score": 0.0}
}
```

## General A/V

```text
# Evaluation Input (General A/V)

You are provided with SOURCE and TARGET multimodal inputs (video and audio).

Source Video Caption:
{source_caption}

Edit Prompt:
{edit_prompt}

Edit Type:
{edit_type}

Checklist subset (general questions only):
{checklist_text}

Instructions:
1. Answer every checklist item above.
2. Keep question_id, question, dimension, subdimension, and modality_tag unchanged.
3. Use both visual and audio evidence.
4. For Instruction Following state questions, target-only evaluation is allowed.
5. Judge each dimension by the rules in the system prompt; for Fidelity, always compare source and target.

Respond in JSON:
{
  "visual_discrepancy_analysis": "short multimodal discrepancy analysis",
  "evaluations": [
    {
      "question_id": "Q01",
      "dimension": "Instruction Following",
      "subdimension": "instruction_correctness",
      "modality_tag": "general",
      "question": "...",
      "observation": "...",
      "answer": "Yes/No",
      "justification": "..."
    }
  ],
  "summary": {"total_questions": {len(checklist_items)}, "yes_count": 0, "no_count": 0, "score": 0.0}
}
```
\end{lstlisting}
\subsection{Realism Evaluation Prompt}
\label{app:prompt-realism-evaluation}

The realism evaluator judges whether the edited target video remains physically and temporally plausible after the requested modification.

\subsubsection{Realism Evaluator}
\label{app:prompt-realism-evaluator}
\begin{lstlisting}[style=promptstyle]
# Realism Evaluation Prompt

```text
# Task
Rate the Realism of the TARGET edited audio-video on 5 sub-dimensions (each 1-5).
Realism = whether the target is natural, well-formed and coherent --- NOT whether it follows the instruction.

# Rules
- Judge the TARGET only. Edit instruction / caption are context, not requirements.
- Do NOT penalize an edit that was not performed; if an attempted change is simply absent, ignore it.
- Audio dims: output "NA" if there is no audio / no clear sound source. Do not default to 5 or 1.
- Motion blur is scene-adaptive: blur from fast motion/camera is fine; blur on static content is a defect.

# Shared 1-5 scale
5 = fully natural & coherent, no issues.
4 = minor: slight warp/distortion/blur, recognition & perception unaffected.
3 = noticeable: looks AI-generated / illogical, or a physics violation, but no glaring error.
2 = clear error: deformation, abnormal/extra object, sudden appearance/disappearance, third hand,
    floating object, clipping-through, or a visibly mismatched/unblended edit region.
1 = severe / broken.

# Sub-dimensions
VISUAL
- object_integrity: each subject's own form & existence. Penalize structural errors (reverse joints,
  rubber limbs, extra/fused fingers, melting faces, rigid bodies bending like jelly, texture morphing)
  and existence MUTATIONS (objects popping in/out without occlusion, flicker, category/identity change).
  Reasonable occlusion/exit is fine.
- interaction_physics: physical & commonsense logic of interactions. Penalize violations of gravity,
  contact/collision, liquids, slicing/breaking/pouring, support (feet on ground), clipping-through,
  and action-object mismatch (e.g. a person on a motorcycle making a pedaling motion).
- naturalness: does it look like a real, coherently-styled video rather than an AI edit. Penalize an
  edited region that does not match the background (lighting, shadow, color, seam), inconsistent art
  style across the frame, an obvious AI-generated look, and warping/flicker/jitter artifacts.
AUDIO
- AAS: signal purity --- clipping, static, dropout, robotic/distorted sound. NA if no audio.
- MTC: does each sound's timbre match the visible material and the room acoustics. NA if no sound source.

# Context (scene understanding only)
Edit instruction: {edit_prompt}
Source caption: {caption}

# Output - return ONLY this JSON:
{{"object_integrity": <1-5>, "interaction_physics": <1-5>, "naturalness": <1-5>,
  "AAS": <1-5 or "NA">, "MTC": <1-5 or "NA">,
  "reason": "<one concise sentence naming the dominant issue, or 'none'>"}}
```
\end{lstlisting}